\documentclass[aps,twocolumn,showkeys,showpacs,preprintnumbers,prd,superscriptaddress,nofootinbib,10pt]{revtex4-1}
\usepackage{graphicx,epsf,bm,amsmath,amsfonts,amssymb,epstopdf,natbib,hyperref,color,verbatim,multirow,bm,appendix}
\hypersetup{colorlinks=true,urlcolor=blue,citecolor=blue,linkcolor=blue,menucolor=blue,anchorcolor=blue,filecolor=blue}

\date{\today}

\begin{document}

\title{Impact of DESI BAO Data on Inflationary Parameters: stability against late-time new physics}

\author{Simony Santos da Costa}
\email{simony.santosdacosta@unitn.it}
\affiliation{Department of Physics, University of Trento, Via Sommarive 14, 38123 Povo (TN), Italy}
\affiliation{Trento Institute for Fundamental Physics and Applications (TIFPA)-INFN, Via Sommarive 14, 38123 Povo (TN), Italy}

\begin{abstract}
In this work, I investigate the impact of Dark Energy Spectroscopic Instrument (DESI) Baryonic Acoustic Oscillations (BAO) data on cosmological parameters, focusing on the inflationary spectral index $n_s$, the amplitude of scalar perturbations $A_s$, and the matter density parameter $\omega_m$. By examining different models of late-time new physics, the inflationary parameters were revealed to be stable when compared with the baseline dataset that used the earlier BAO data from the SDSS collaboration. 
When combined with Cosmic Microwave Background (CMB) and type Ia supernovae (SNeIa), DESI BAO data leads to a slight reduction in $\omega_m$ (less than 2\%) and modest changes in $A_s$ and $n_s$, if compared with the same combination but using SDSS BAO data instead, suggesting a subtle shift in matter clustering. These effects may be attributed to a higher expansion rate from dynamical dark energy, changes in the recombination period, or modifications to the matter-radiation equality time. Further analyses of models with dynamical dark energy and free curvature show a consistent trend of reduced $\omega_m$, accompanied by slight increases in both $n_s$ and $H_0$. The results emphasize the importance of the DESI BAO data in refining cosmological parameter estimates and highlight the stability of inflationary parameters across different late-time cosmological models.
\end{abstract}

\maketitle
\newpage

\section{Introduction}
\label{sec:introduction}

The $\Lambda$CDM model is the most successful in describing the cosmological and astrophysical observations on different redshifts and scales, including observations of the cosmic microwave background, clustering of galaxies, type Ia supernovae, and Big Bang Nucleosynthesis (BBN)~\cite{SupernovaSearchTeam:1998fmf,SupernovaCosmologyProject:1998vns,DES:2017qwj,Planck:2018vyg,SPT:2019fqo,ACT:2020gnv,eBOSS:2020yzd,KiDS:2020suj,Mossa:2020gjc,Brout:2022vxf}. In this context, $\Lambda$ stands for an exotic component of dark energy (DE), represented by a cosmological constant, driving the current accelerated expansion and responsible for $\sim 70\%$ of the energy content of the Universe, CDM stands for Cold Dark Matter, accounting for other $\sim25\%$ of the energy content, and the ordinary matter and radiation account for the energy budget's last $\sim 5\%$. However, we still have no conclusive knowledge about the nature of the dark sector, which remains one of the most profound mysteries in cosmology.

To pose $\Lambda$CDM as the concordance cosmological model, we need an extra ingredient, cosmic inflation, a period of rapid, exponential expansion in the universe’s first moments. Inflation was initially postulated to address inconsistencies in the standard Big Bang model, including the large-scale uniformity of the universe and the origins of its current structure. 
It predicts distinct imprints on the CMB, the large-scale distribution of galaxies, and other cosmological observables. The origin for these features emerges due to primordial quantum fluctuations inducing curvature perturbations $\mathcal{R}$, which relates to fluctuations in the matter density field $\delta$. The result is a nearly scale-invariant power spectrum~\cite{Mukhanov:1981xt,Mukhanov:1982nu,Hawking:1982cz,Starobinsky:1982ee,Guth:1982ec,Bardeen:1983qw}, that in the concordance model is typically parametrized as a power-law $P(k)=A_s(k/k_{*})^{n_s-1}$, where $A_s$ is its amplitude, $k$ is the wave-number of a perturbation, and $n_s$ is the scalar spectral index.

Inflation also predicts primordial gravitational waves, originating from the tensor perturbations, which are characterized through the tensor-to-scalar ratio parameter $r$. The parameters $n_s$ and $r$ encode fundamental information about the inflationary dynamics and are crucial for distinguishing between competing inflationary models. Indeed, we can find hundreds of models to describe inflation in the literature~\cite{Martin:2013tda}. The simplest models include a single scalar field with minimal kinetic terms, slowly rolling down its potential. The Friedmann-Lemaître and Klein-Gordon equations dictate the dynamic and is well described via the slow-roll parameters in terms of the derivatives of the potential. In this way, it is possible to relate the parameters $n_s$ and $r$ with the field dynamics and study the model's viability confronting the theoretical predictions with observational data (see \cite{Martin:2013tda} which applied this methodology for an extensive list of models).

Over the past few decades, we have seen remarkable advancements in observational technology, including a significant accumulation of data from various ongoing and upcoming surveys. Notably, Stage IV surveys, such as the DESI and Euclid, along with several upcoming CMB Stage IV missions, are expected to greatly enhance our understanding of the universe. 
Indeed, the recent DESI Year 1 BAO results brought intriguing hints of a dynamic, time-evolving Dark Energy component~\cite{DESI:2024mwx}. By capturing the spectra of millions of galaxies, DESI also promises to refine the constraints on a range of cosmological parameters, such as the matter density in the universe, spatial curvature, neutrino mass, and the amplitude of primordial fluctuations.
The results from DESI BAO data are particularly important because allow us to investigate the matter content information present in galaxy clustering, which brings imprinted in itself a preferred scale, the sound horizon at the baryon drag epoch, $r_d$, determined by physics around recombination and before. Further, since this feature is stretched due to the universe's expansion, measurements of the BAO scale can indirectly provide information about the inflationary epoch and directly about the late-time universe. As will be better explained later, uncalibrated BAO measurements can constrain $\Omega_m$, which is correlated with the inflationary parameters.

Another interesting point about the results from Stage IV surveys consists in bringing some insights about important inconsistencies that have been suggested within the $\Lambda$CDM model. Undoubtedly, the most prominent of these tensions is the Hubble tension, accompanied by a milder tension in \( S_8 \) (for recent reviews, see \cite{Perivolaropoulos:2021jda,Abdalla:2022yfr}). These discrepancies may indicate potential shortcomings in the concordance \( \Lambda \)CDM cosmological model~\cite{Akarsu:2024qiq} and possibly the necessity for new physics. Two main avenues have been taken to address the $H_0$ tension: either modifications before recombination or in the late-time universe. The first one usually tries to reduce the sound horizon at recombination in the way that $H_0$ increases. 
The second route of including new physics in the late Universe, for example, through models of dynamic dark energy, is now well known to be less effective once the fits to BAO and SNeIa are worsened due to $r_d$ not being altered~\cite{Bernal:2016gxb,Addison:2017fdm,Lemos:2018smw,Aylor:2018drw,Knox:2019rjx,Arendse:2019hev,Efstathiou:2021ocp,Cai:2021weh,Keeley:2022ojz}. The interested reader could check \cite{Vagnozzi:2023nrq} to find an excellent discussion on the promising scenario of addressing the Hubble tension by combining early- and late-time new physics.

At this stage, one could think that having early- and late-time new physics acting on different epochs would not affect each other. However, as noticed above, altering the physics before recombination to produce a smaller $r_d$ will modify today's expansion rate, bringing different probes to agreement when talking about the $H_0$ value. What about the other way around? Could the indications of new physics in the late universe obtained by DESI BAO data indirectly impact the inference of parameters related to the primordial universe, such as the inflationary parameters?
To address this question, in this study, I integrate the latest DESI BAO measurements with CMB and Type Ia Supernovae data to revisit key cosmological constraints, particularly regarding the inflationary parameters $n_s$ and $r$. Additionally, I explore the impact of these new measurements on various models that modify the $\Lambda$CDM model at late-times.%

The structure of this paper is as follows: in Sec.\eqref{sec:degeneracyeffects} I discuss the main role of BAO along the universe's expansion history and the primary effects on the temperature power spectrum due to additional correlations arising from
changes in $r_d$ and $H_0$. In Sec.\eqref{sec:dataandmethod}, I present the datasets and the methodology to perform the analyses. In Sec.\eqref{sec:results}
 I discuss the main results and in Sec.\eqref{sec:conclusions}, the main conclusions of this work.

\section{The role of BAO to understand the universe's expansion history}
\label{sec:degeneracyeffects}

The Baryonic Acoustic Oscillations serve as a ``standard ruler" since it is encoded in the clustering of matter through a preferred scale, the sound horizon at the drag epoch of the early universe\footnote{The distance that sound waves could travel between the reheating and the time when the baryons decoupled from photons.}~\cite{DESI:2024mwx}:
\begin{eqnarray}
    r_d = \int_{z_d}^{\infty}\frac{c_s(z)}{H(z)}dz,
\end{eqnarray}
which is driven by physics around recombination and earlier. Thus,  BAO allows precise measurements of the universe's expansion history and provides a powerful method for constraining cosmological parameters.  
In the $\Lambda$CDM model, the expansion rate $H(z)$ is given in terms of the Hubble constant and the density parameters' evolution as
\begin{equation}
    H(z) = H_0 \sqrt{\Omega_m(1+z)^3+\Omega_r(1+z)^4+\Omega_\Lambda},
\end{equation}
where $\Omega_m$, $\Omega_r$, and $\Omega_\Lambda$ are the energy densities relative to critical in matter, radiation, and the cosmological constant, respectively. 
Along the universe's expansion, the BAO feature appears as a comoving galaxy separation of $ r_d \sim 150$Mpc. Notice, however, that for a given galaxy distribution at a given redshift $z$ with a preferred angular separation $\Delta\theta$, when measuring the comoving distance to that redshift, it is equivalent to measuring a distance $D_M(z)\equiv r_d/\Delta\theta$ for pairs of galaxies perpendicular to the observer's line-of-sight. If, instead, the separation vector is aligned with the line-of-sight, we observe a preferred redshift separation $\Delta z$, such that measuring the comoving distance interval gives us the Hubble parameter at that redshift through $D_H\equiv c/H(z)=r_d/\Delta z$.  A particular case happens when we have an angle-averaged distance along, and perpendicular to, the line of sight to the observer, given by $D_V\equiv(zD_M(z)^2D_H(z))^{1/3}$. In all of these cases, BAO alone puts constraints on $D_M(z)/r_d$, $D_H(z)/r_d$, and $D_V(z)/r_d$. Therefore, unless one calibrates $r_d$ through a recombination model, one can only get constraints for the matter density $\Omega_m$ and the combination parameter $r_dH_0$ when using BAO measurements.

The direct correlation of BAO with the matter density distribution is remarkable. Looking further, we can observe its interplay with other cosmological parameters when observing this feature in the galaxy correlation function as well as wiggles on the matter power spectrum. The latter is related to the autocorrelation function of the density contrast, $\delta\equiv (\rho(x)-\bar{\rho})/\bar{\rho}$, in the Fourier space as
\begin{eqnarray}
    P(k)\equiv|\delta^2(k)|.
\end{eqnarray}
In general, the form of the spectrum will depend on how the amplitude of a fluctuation of fixed comoving wave-number $k$ (or wavelength $\lambda$) grows with time, which in turn depends on which constituent is dominant along the universe's expansion\footnote{The growth of each constituent is well described by relativistic perturbation theory.}.

In the $\Lambda$CDM context, the matter density fluctuations are related to the primordial quantum perturbations described by the inflationary scenario, which, as discussed earlier, can be well approximated by the power law:
\begin{eqnarray}
    P(k)= A_s \left(\frac{k}{k_\star}\right)^{n_s-1}.
\end{eqnarray}
Therefore, the BAO feature can put indirect constraints on inflationary physics by relating the restrictions on $\Omega_m$ with its correlation to $n_s$ and $A_s$. 
Indeed, interesting studies addressing the Hubble tension, for example, have found important correlations between the parameters of the $\Lambda$CDM model, especially when incorporating BAO measurements~\cite{Lin:2020jcb,McDonough:2023qcu,Simon:2023hlp,Pedrotti:2024kpn}. Previous BAO measurements from the Baryon Oscillation Spectroscopic Survey (BOSS)~\cite{BOSS:2016wmc} and the extended BOSS (eBOSS) collaboration~\cite{eBOSS:2020yzd}, along with other observational data, have been extensively used to constraint the standard six-parameters of the $\Lambda$CDM, i.e. the cold dark matter density $\omega_{c}=\Omega_c h^2$,~\footnote{We are using the reduced Hubble constant $h$ defined as $h\equiv H_0/(100$ km/s/Mpc$ )$.} the physical baryon density $\omega_b=\Omega_b h^2$, the acoustic angular scale $\theta$, the amplitude of the primordial scalar power spectrum $A_s$ and its spectral index $n_s$, and the re-ionization optical depth $\tau$, and also beyond $\Lambda$CDM models, given its complementary information to that from type Ia supernovae and CMB observations.

The advent of DESI as the first Stage IV survey in operation opens the possibility of improving the cosmological constraints by increasing the sample size by a factor of 10 if compared with previous BOSS and eBOSS~\cite{DESI:2024mwx}. This is particularly important because it will allow for tight constraints on matter density, the equation of state of dark energy, spatial curvature, and the amplitude of primordial fluctuations. 
The release of the first year of DESI BAO data suggests deviations from the $\Lambda$CDM model through dynamical dark energy, which led to the development of various works studying the implications of these results, ranging from studies approaching the Hubble tension~\cite{Jiang:2023bsz,Jiang:2024nha,Wang:2024dka,Lynch:2024hzh} (mainly due to previous studies hinted phantom models as a possible way to solve it~\cite{DiValentino:2016hlg,Vagnozzi:2019ezj}), the nature of dark energy (including studies of quintessential models~\cite{Tada:2024znt,Ramadan:2024kmn,Bhattacharya:2024hep} and interacting DE models~\cite{Giare:2024smz,Giare:2024ocw}), neutrino's masses~\cite{Jiang:2024viw,Naredo-Tuero:2024sgf}, modified gravity~\cite{Chudaykin:2024gol}, etc.
Some of these studies corroborated the previous correlations between the cosmological parameters when addressing the Hubble tension. 
In this section, I summarize the main synergies between those cosmological parameters and discuss how they open up new possibilities for refining our understanding of the universe's expansion history and the fundamental parameters that govern it.

The reason behind the correlation between the sound horizon at recombination, $r_\star$, and the Hubble constant is because the first when associated with the distance from us up to the recombination epoch, $D(z)$, establishes an angular size, $\theta_\star\equiv r_\star/D(z_\star)$, such that to keep this angle fixed when diminishing $r_\star$, one needs to increase $H_0$. The size of the angular scale $\theta_\star$ is well determined through CMB observations, using the location of the acoustic peaks and their spacing, and also through BAO using the two-point correlation function in the distribution of galaxies. Therefore, it is clear that the predictions of including new physics before recombination would alter the expansion history at late times and, thereby, could be tested using CMB, BAO, and SNe Ia datasets.

\begin{figure*}
    \centering
    \includegraphics[width=1\linewidth]{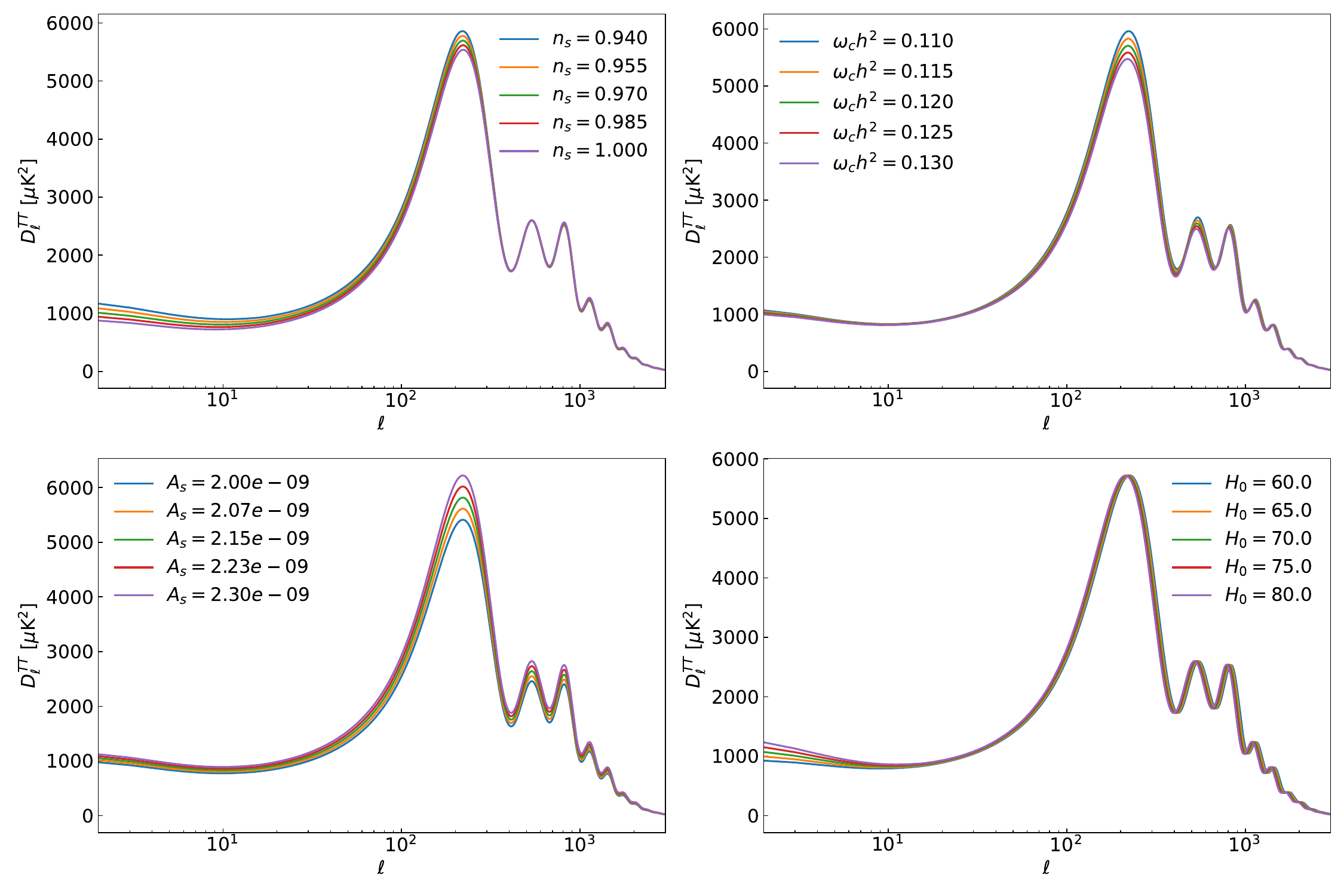}
    \caption{Effects of changing the parameters $n_s$, $\omega_ch^2$, $A_s$, and $H_0$ on the CMB temperature fluctuations power spectrum. In each panel, I fixed all the cosmological parameters (including $H_0$) and varied only the one indicated on the legend. Note how they are connected. To recover the same amplitude on the first peak after increasing $n_s$, it is necessary to decrease $\omega_ch^2$ or change $A_s$. However, changing $\omega_ch^2$ could produce a shift on the second peak position, which could be recovered by changing $H_0$.}
    \label{fig:effects}
\end{figure*}

The interplay of additional correlations arising from changes in \( r_d \) and \( H_0 \) is depicted in Fig.\eqref{fig:effects}, where I highlight the primary effects on the temperature power spectrum. Notably, modifications to \( H_0 \) (shown in the bottom panel on the right) result in shifts in the positions of the peaks as well as changes in the slope at low multipoles. Under the slow-roll approximation, the power spectrum is connected to the Hubble parameter during inflation (\( P(k) \propto H_{inf}^2/\epsilon \)). Therefore, an increase in \( H_0 \) leads to a corresponding rise in the spectral index \( n_s \), which necessitates an adjustment in \( A_s \) to restore the ``correct” amplitude (shown in the bottom left panel). This adjustment can also be achieved by decreasing the matter density parameter. As demonstrated in the top right panel, a reduction in cold dark matter increases the power spectrum's amplitude, similar to the effect of increasing \( A_s \). Consequently, modifications to the physics surrounding recombination, as tested through BAO observations, may yield valuable insights into the inflationary phase by influencing the parameters \( A_s \) and \( n_s \). The underlying mechanism is associated with the reduction of the sound horizon, which causes perturbations to enter the horizon later, resulting in a suppression of power at larger scales (low multipoles).

In the subsequent sections, I will comprehensively examine the valuable insights afforded by DESI BAO data regarding the early universe.

\section{Data and Methodology}
\label{sec:dataandmethod}

As previously discussed, I will use the DESI BAO year 1 release dataset to investigate potential effects associated with the primordial universe. In conjunction with the BAO data, I will also integrate external datasets to enrich our investigation. The datasets are described below:
\begin{itemize}
    \item \textbf{Baryonic Acoustic Oscillations - SDSS:} The earlier BAO dataset composed by the 6dFGS~\cite{Beutler:2011hx}, SDSS-MGS~\cite{Ross:2014qpa}, and the eBOSS-DR16~\cite{eBOSS:2020yzd} measurements.
    \item \textbf{Baryonic Acoustic Oscillations - DESI:} The full DESI BAO sample, including clustering of galaxies of the bright galaxy survey (BGS), luminous red galaxies (LRG), emission line galaxies (ELG), quasars and Lyman$-\alpha$ forest, which provides information on $D_H/r_d$, $D_M/r_d$, and $D_V/r_d$, in the redshift range of $0.1<z<4.2$~\cite{DESI:2024mwx}.
    \item \textbf{Type Ia Supernovae:} Distance moduli measurements in the range $0.001<z<2.26$ of 1550 spectroscopically-confirmed SNeIa from Pantheon+ compilation~\cite{Scolnic:2021amr,Peterson:2021hel} (hereafter denoted as PP); 2087 SNeIa in the range $0.010<z<2.26$ from Union3 compilation~\cite{Rubin:2023ovl}; and 1829 SNe Ia in the range $0.025<z<1.3$ from Dark Energy Survey Year 5 data release~\cite{DES:2024jxu} (from now on denoted as DESY5).
       \item \textbf{Cosmic Microwave Background:} I consider the most recent release of Planck maps (PR4) which utilizes the NPIPE code~\cite{Planck:2020olo}. For the high-$\ell$ multipoles, I apply the hillipop likelihood~\cite{Couchot:2016vaq}, while the lollipop~\cite{Tristram:2020wbi} serves as the likelihood for low-$\ell$ polarization. Regarding the low-$\ell$ TT power spectrum, I utilize the publicly available commander likelihood~\cite{Planck:2019nip}. For the CMB lensing, I use the Planck PR4 likelihood derived from the temperature 4-point function~\cite{Carron:2022eyg} and the ACT DR6 lensing power spectrum likelihood~\cite{ACT:2023kun}. Lastly, the latest BICEP/Keck likelihood on the BB power spectrum is also incorporated~\cite{BICEP:2021xfz}.
\end{itemize}
Further, I combined CMB and SNeIa from the Pantheon+ sample as the baseline dataset. To examine the main impact of BAO measurements on the constraints, I combined the baseline with the earlier BAO sample composed of the measurements taken by SDSS and eBOSS collaborations, denoted as baseline+SDSS, and a second combination replacing this BAO data with the DESI BAO dataset, denoted as baseline+DESI. There will be cases where I also replace the SNeIa from Pantheon+ with Union3 or DESY5 samples.

Considering the suggestive indication of deviations from the standard model with an evolving dark energy component found by DESI collaboration, I decided to work with different scenarios of evolving dark energy models attempting to catch the same feature and also to understand if late-time new physics implies in any modification for the predictions of the inflationary parameters. In addition to evolving dark energy models, I also consider the $\Lambda$CDM model with free spatial curvature\footnote{I choose this model based on the possible preference for a spatially closed Universe from Planck
data~\cite{DiValentino:2019qzk}, see also Refs.~\cite{Handley:2019tkm,Efstathiou:2020wem,DiValentino:2020hov,Benisty:2020otr,Vagnozzi:2020rcz,Vagnozzi:2020dfn,Yang:2021hxg,Cao:2021ldv,Gonzalez:2021ojp,Dinda:2021ffa,Zuckerman:2021kgm,Bargiacchi:2021hdp,Glanville:2022xes,Bel:2022iuf,Yang:2022kho,Stevens:2022evv,Favale:2023lnp,Qi:2023oxv} for recent discussions.} and the flat $\Lambda$CDM as the reference model.

A straightforward model of evolving dark energy is the recognized  Chevallier-Polarski-Linder (CPL) parametrization $w(z)=w_0+w_a z/(1+z)$~\cite{Chevallier:2000qy,Linder:2002et}, which reduces to the $\Lambda$CDM for $w_0=-1$ and $w_a=0$. This model is the simplest extension for a dark energy component changing with time, and distinguishing it from a cosmological constant is a key point to understanding the nature of dark energy. I also consider a particular case of this parametrization where holds the limit $w(z)>-1$, which avoids the phantom scenarios~\cite{Vagnozzi:2018jhn}. I denote these two CPL cases as $\omega_0 \omega_a$CDM and wzlg-1, respectively. Lastly, I consider the interesting scenario of a sign switching cosmological constant, $\Lambda_s$, recently studied in \cite{Akarsu:2021fol, Akarsu:2022typ,Toda:2024ncp} as an attempt to alleviate the Hubble tension, and which is based on a transition from an anti-de Sitter vacuum to a de Sitter vacuum at a certain redshift $z_\dag$.

The priors considered for the standard six cosmological parameters and the additional model parameters are displayed in Tab.~\eqref{tab:priors}. 
The theoretical predictions for the background expansion and for the CMB power spectra were derived using the \texttt{CAMB} code~\cite{Lewis:1999bs}. To sample the parameter space, I use the Monte Carlo Markov Chain (MCMC) method as implemented in the \texttt{Cobaya} package~\cite{Torrado:2020dgo}. The final chains are considered to be converged when achieving the Gelman-Rubin parameter $R-1<0.02$~\cite{Gelman:1992zz}. To process the chains and analyze the results I used the \texttt{GetDist} package~\cite{Lewis:2019xzd}.

\begin{table}
    \centering
    \begin{tabular}{l cccc}
    \hline
    \hline
      type  & parameters & default & prior\\
       \hline 
       standard  & $\omega_b$ & $-$  & $[0.005,0.1]$\\
         & $\omega_c$ & $-$  & $[0.001,0.99]$\\
         & $100\theta$ & $-$ & $[0.5,10]$\\
         & $\tau$ & $-$ & $[0.01,0.8]$\\
         & $\ln(10^{10} A_s)$ & $-$ & $[1.61,3.91]$\\
         & $n_s$ & $-$ & $[0.8,1.2]$\\
      \hline 
      tensor perturbations   & $r$ & $-$ & $[0,3]$\\
      \hline
      background   & $H_0$ [Km/s/Mpc] & $-$ & $[20,100]$\\
         \hline
      adding free-curvature  & $\Omega_k$ & 0 & $[-0.3,0.3]$\\
      \hline 
      sign switching $\Lambda_s$ & $z_\dag$ & $-$ & $[1,3]$\\
      \hline
      $\omega_0\omega_a$CDM   & $w_0$ & $-1$ & $[-3,1]$\\
            & $w_a$ & $0$  & $[-3,2]$\\
      \hline
      wzlg-1 & $w_0$ & $-1$ & $[-1,1]$\\
                     & $w_a$ & $0$  & $[-1,2]$\\
        \hline
    \end{tabular}
    \caption{Uniform priors for the parameters sampled with the MCMC analyses. The first eight lines show the parameters common to all models. Next, one can see the parameters representing the main features of the extended models: curvature, the redshift of the moment when the cosmological constant changes sign, and the parameters characterizing the parametrization for dynamical dark energy.}
    \label{tab:priors}
\end{table}

\section{Results and Discussion}
\label{sec:results}

\begin{figure*}
    \centering
    \includegraphics[width=.32\linewidth]{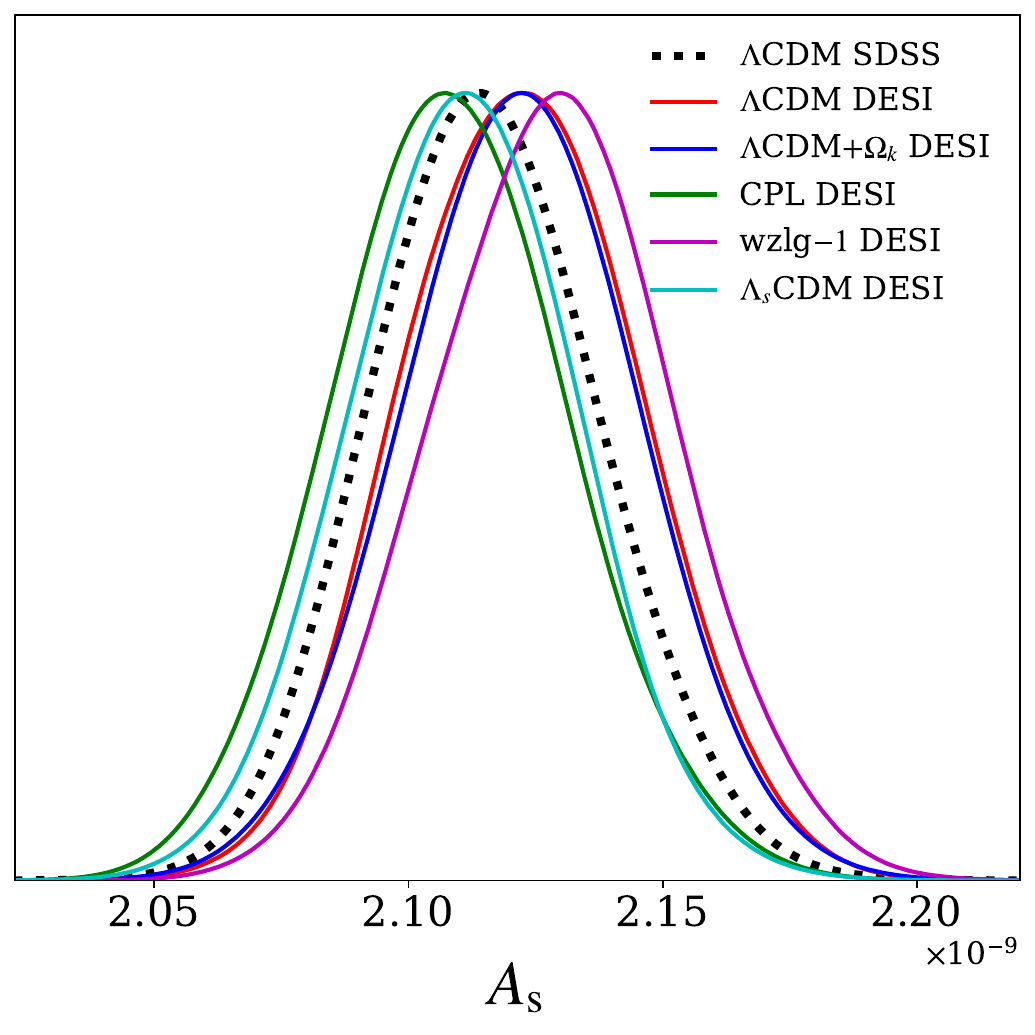}
    \includegraphics[width=.32\linewidth]{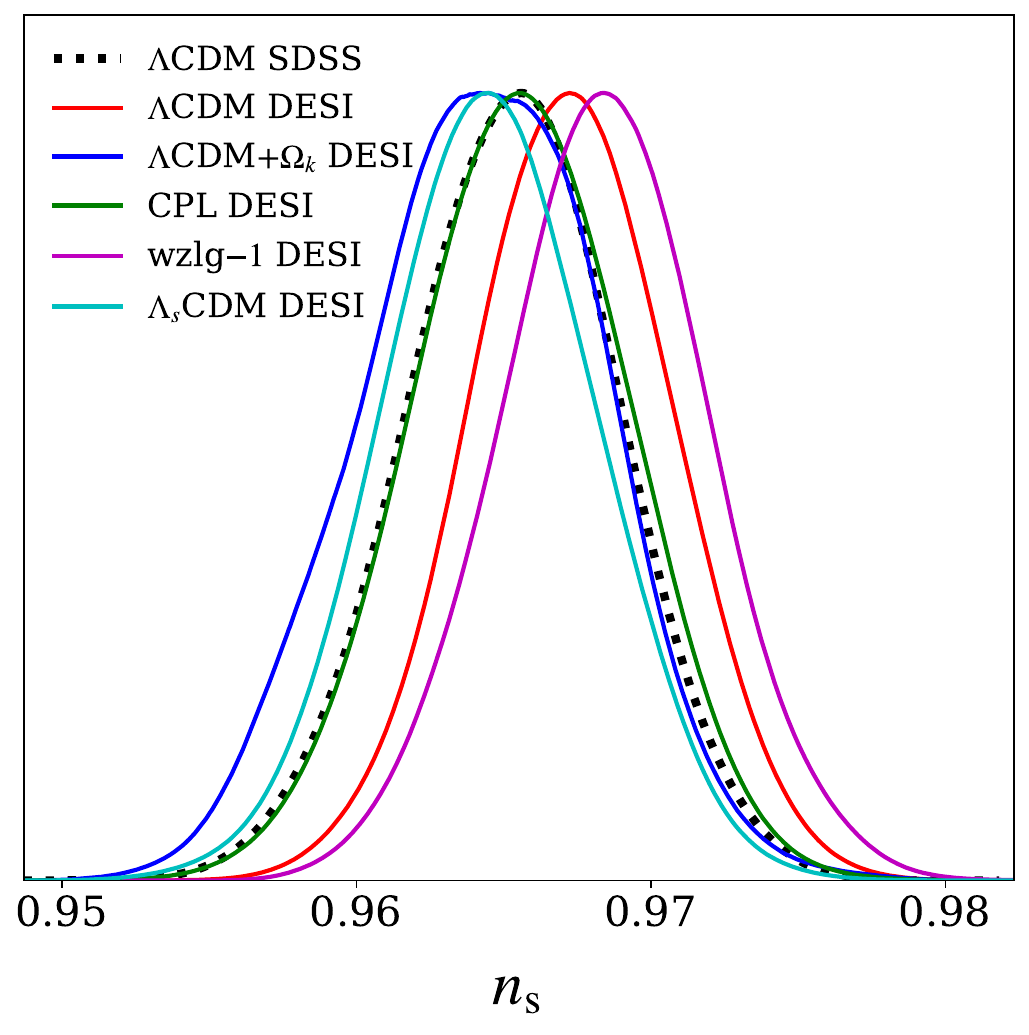}
    \includegraphics[width=.32\linewidth]{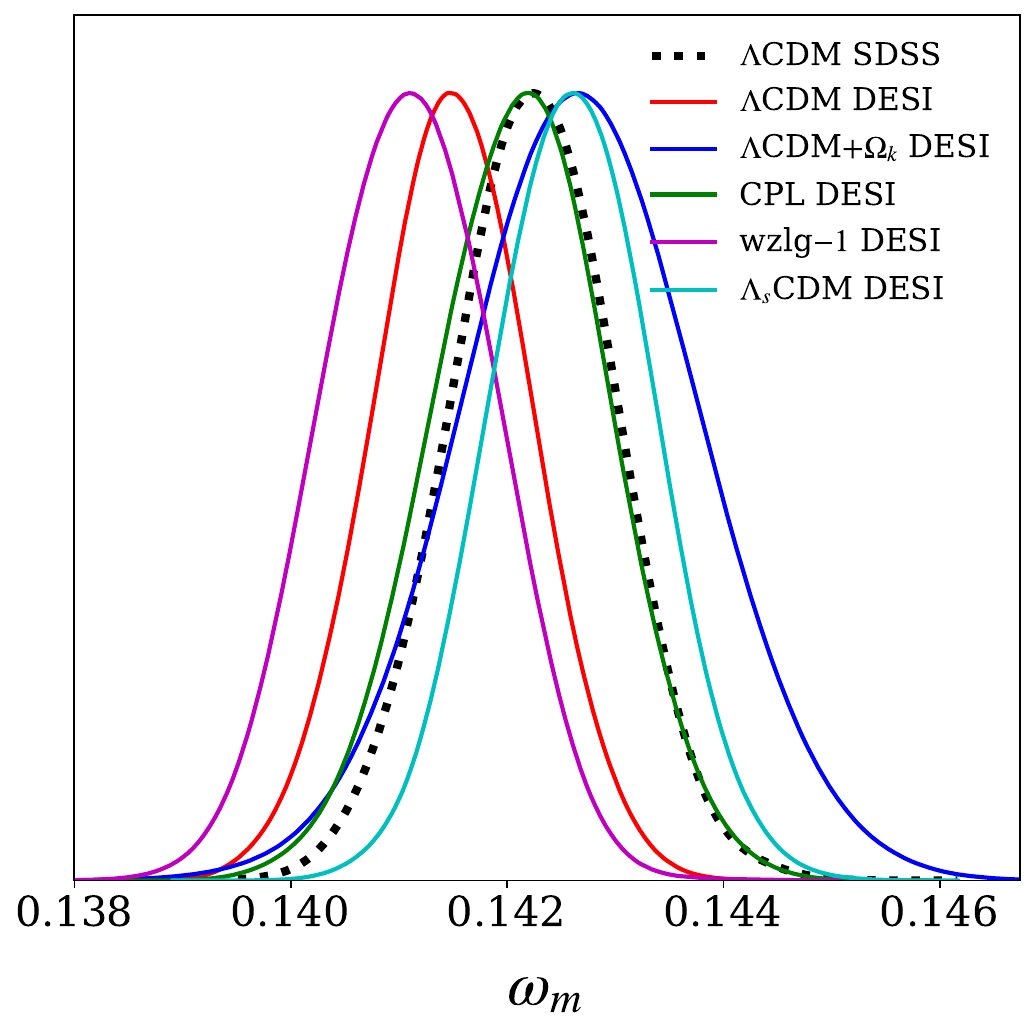}
    \caption{The one-dimensional marginalized posterior distribution for the most correlated parameters, considering the baseline dataset combination CMB+SDSS+PP for the $\Lambda$CDM model (black dotted line), and the dataset combination CMB+DESI+PP for the $\Lambda$CDM model (red solid line), the $\Lambda$CDM with free curvature (blue solid line), CPL parametrization with (magenta line) and without (green line) the $w(z)>-1$ constrain, and lastly the sign-switching cosmological constant (cyan line).}
    \label{fig:posteriors}
\end{figure*}

\begin{figure*}
    \centering
    \includegraphics[width=1\linewidth]{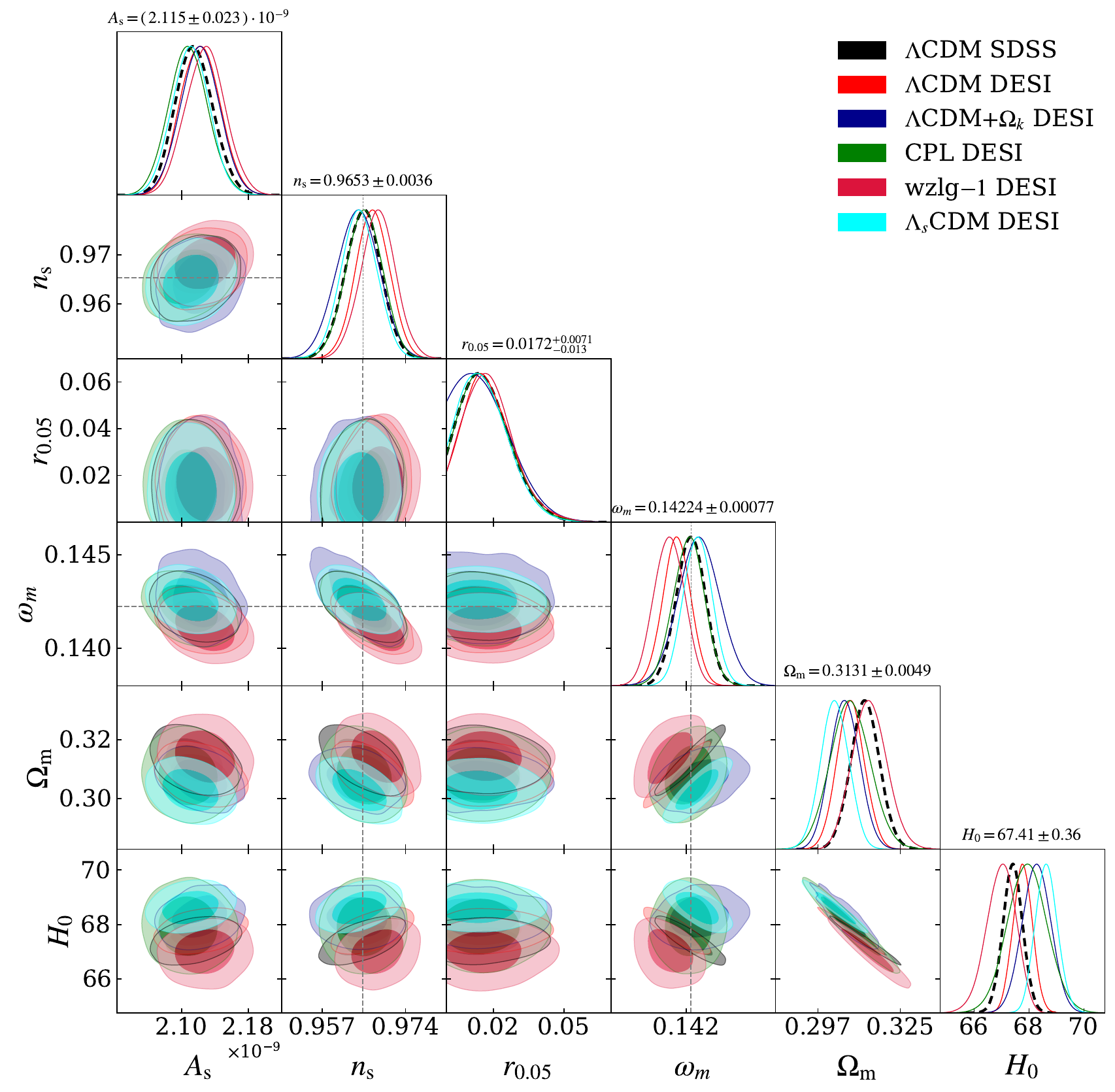}
    \caption{The two-dimensional and one-dimensional posterior probability distributions for the most correlated parameters, considering the baseline dataset combination CMB+SDSS+PP for the $\Lambda$CDM model (black dotted line and black contours), and the dataset combination CMB+DESI+PP for the $\Lambda$CDM model (redlines and contours), the $\Lambda$CDM with free curvature (blue lines and contours), CPL parametrization with (magenta lines and contours) and without (green lines and contours) the $w(z)>-1$ constrain, and lastly the sign-switching cosmological constant (cyan lines and contours). Note the stability on the $n_s - r$ plane and how it is associated with the simultaneous variations on $\Omega_m$ and $H_0$. Note also how all the models are consistent within $1\sigma$ C.L. with the baseline combination, with the most pronounced shifts happening for the physical matter density $\omega_m$ for the CPL parametrization, avoiding the phantom regime.}
    \label{fig:triplot}
\end{figure*}

The main results of our analyses are shown in Table \eqref{tab:results}, and Figs. \eqref{fig:posteriors} and \eqref{fig:triplot}. In Table \eqref{tab:results}, I highlight the main parameters of interest: the ones presenting the strongest correlation with the inflationary parameter, $n_s$, and the ones characterizing the extended models.
Upon preliminary analysis, it is clear that the inferred parameters across the various models examined are consistent with the flat $\Lambda$CDM model using the baseline+DESI dataset, remaining within a $1\sigma$ confidence level (CL).

Notably, the BAO data from DESI moderately impacts the matter density parameter, which undergoes a marginal reduction of no more than $2\%$ ($0.75\sigma$). In Figure \eqref{fig:posteriors}, I present the posterior distributions for three critical parameters: the amplitude of scalar perturbations ($A_s$), the spectral index ($n_s$), and the physical matter density parameter ($\omega_m$). The effects observed on the parameters $A_s$ and $n_s$ are relatively subdued, with only slight shifts in their central values—each exhibiting an increment of less than $1\%$ ($0.4\sigma$). This observation can be further elucidated by examining the discussions in Section \eqref{sec:degeneracyeffects}, which illuminate an anti-correlation between $A_s$ and $n_s$ with the matter density parameter.

To further understand these dynamics, it is valuable to interpret the implications of the DESI BAO data, which indicates a slight decrease in the clustering of matter. This observational trend may arise from various factors, such as late-time effects linked to a higher expansion rate stemming from dynamical dark energy or a recalibration of the early Integrated Sachs-Wolfe (eISW) amplitude to agree with the CMB data. Since I am not considering modifications to physics during recombination and are not affecting the onset of radiation-matter equality, the duration of gravitational potential decay is not altered. Thus, the eISW contribution does not change. The eISW effect is closely associated with the height of the first acoustic peak and demonstrates strong consistency with the CMB Planck data~\cite{Hou:2011ec,Kable:2020hcw,Vagnozzi:2021gjh}. The total matter density is another factor impacting the amplitude of the CMB temperature power spectrum. Consequently, the eISW effect constrains the extent to which this parameter can vary, with minor deviations offset by adjustments in the related parameters.

In light of the evolving dark energy hints identified by the DESI collaboration, I investigated late-time scenarios that accommodate dynamic dark energy alongside the $\Lambda$CDM model with free curvature. Our objective was to determine whether the observed trend of a slight decrease in $\omega_m$ and an increase in $n_s$ persists across these models.
Table \eqref{tab:results} presents the findings for the sign-switching cosmological constant model ($\Lambda_s$CDM), the CPL model (both with and without the constraint of $w(z) > -1$), and the $\Lambda$CDM model with free curvature. A consistent pattern of reduced $\Omega_m$ emerges across all models, accompanied by slightly rising values of both $n_s$ and $H_0$. Nonetheless, independent of the late-time modifications considered, the combination of BAO with CMB and SNeIa data always prefers low values of $H_0\sim 66 - 68$ km/s/Mpc (in comparison with the local measurements). These results reflect the stability or a slight increase (decrease) for the inflationary parameter $n_s$ ($r$) due to the simultaneous shifts on $\Omega_m$ and $H_0$.
One can see these features clearly in Fig.\eqref{fig:triplot}, where we observe the stability on the $n_s - r$ plane and the small variations on the contour regions of $\Omega_m$ and $H_0$.

A subsequent effect of the observed stability on the inflationary parameters due to simultaneous shifts in other parameters could be understood as the multidimensionality of the Hubble tension, which was recently studied in Refs.\cite{Pedrotti:2024kpn,Poulin:2024ken}. The authors argue that if both the matter density parameter ($\Omega_m$) and the baryon density parameter ($\omega_b$) are calibrated (through BAO and/or uncalibrated SNeIa and BBN), an increase in the Hubble constant ($H_0$) must be necessarily accompanied by a rise in cold dark matter density ($\omega_c$), with these shifts arising exclusively from late-time expansion history constraints. Note that for this study case, $H_0$ remains close to the lower central values and so $\Omega_m$, resulting in the stability of $n_s$. However, if one considers, for instance, models that alter early-time physics, resulting in an increase in \( H_0 \), we would observe a higher cold dark matter density (\( \omega_c \)) accompanied by elevated values of \( n_s \) to counterbalance the excess of eISW effect. A notable illustration of this situation in actual data is presented in the recent study by Wang et al. (2024)~\cite{Wang:2024rjd}. In their work, the authors integrated pre-recombination solutions, such as Early Dark Energy, with late-time dynamical dark energy to address the Hubble tension, yielding higher values for \( H_0 \) and \( n_s \).

\begin{figure*}
    \centering
    \includegraphics[width=1\linewidth]{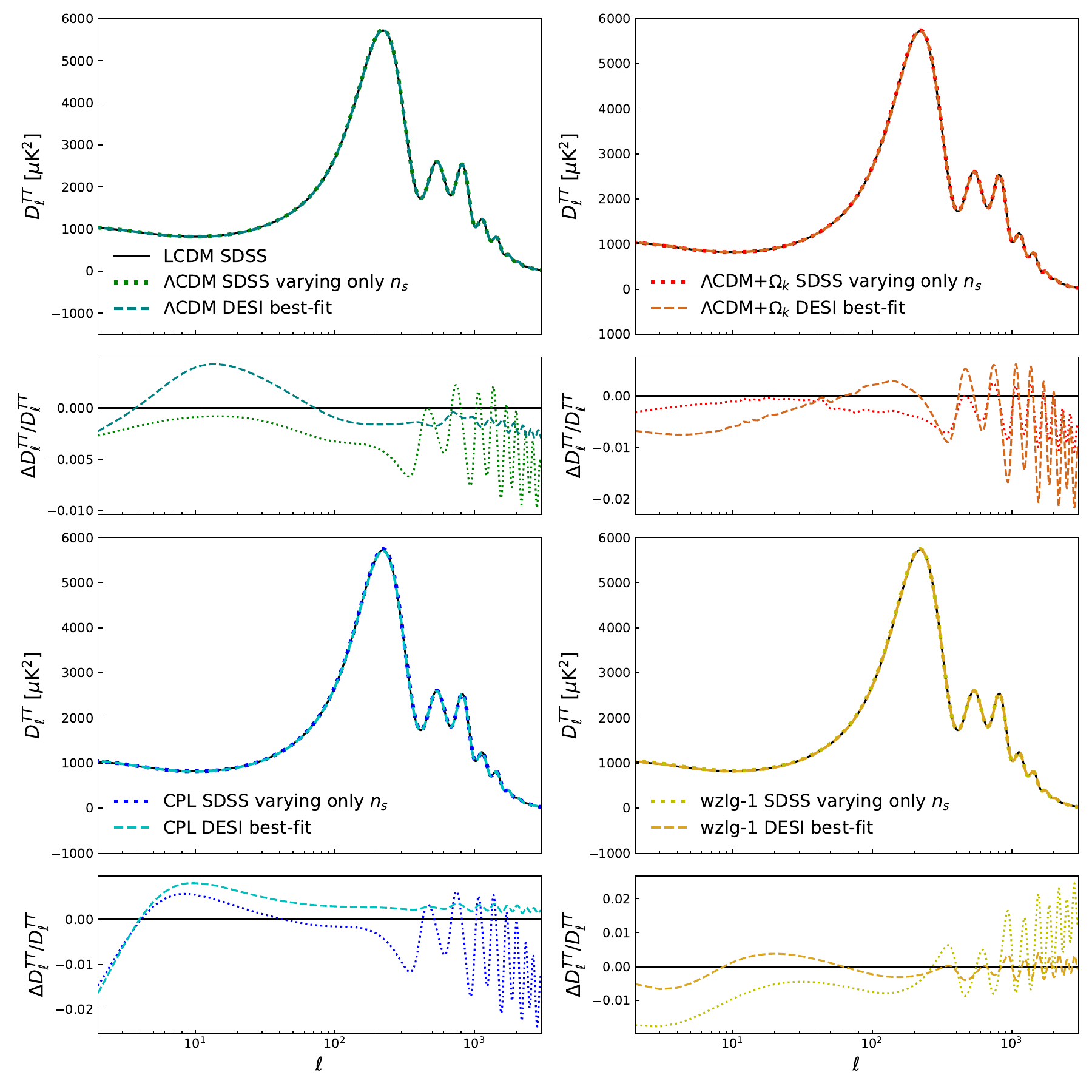}
    \caption{Effects of allowing only $\omega_m$ variations to the CMB temperature anisotropies power spectrum (dotted curves) and considering the best-fit values obtained from the dataset combination CMB+DESI+PP (dashed curves). Observe how the effects on small scales are readjusted to fit the data by the variation on other parameters than $\omega_m$ only.}
    \label{fig:final}
\end{figure*}

Ultimately, the shift in $n_s$ could be interpreted as a change in the slope of the primordial power spectrum of scalar perturbations due to a suppression of growth on small scales (large $\ell$). Indeed, we see an indication of this effect in Fig.\eqref{fig:final}, whereby changing $\omega_m$ and fixing all the other parameters, the main impact happens to be on small scales. Of course, since we are considering different late-time models, we also note the influence of dark energy on low multipoles. After readjusting all the parameter values to fit CMB+DESI+PP data, we bring all the models to a better agreement with $\Lambda$CDM using the CMB+SDSS+PP dataset. 

Finally, I also examined the impact of different supernova samples on the various models analyzed, specifically the flat $\Lambda$CDM model and the CPL model, both with and without the constraint of $w(z) > -1$. Our findings were consistent with those presented in DESI's original paper~\cite{DESI:2024mwx}; the results varied based on the sample utilized, often indicating a more significant deviation from the cosmological constant, with the data favoring a dynamic dark energy model. Nevertheless, all three SNeIa samples agree on the current constraints. This suggests that the primary influence stems from the BAO data, especially since I analyzed all three SNeIa samples in conjunction with the CMB+SDSS data.

\begin{table*}
    \centering
    \scalebox{0.9}{
    \begin{tabular}{l ccccccccc}
    \hline
    \hline
    &  &  &  &  &  &  &  & \\
    model/dataset     & $\Omega_m$ & $H_0$  & $10^3 \Omega_K$ & $\omega_0$ & $\omega_a$ & $n_s$ & $10^2r$ & $z^{\dag}$\\
    &  & [km s$^{-1}$ Mpc$^{-1}$] &  &  &  &  &  & \\ [2mm]
    \hline
     \textbf{Flat $\Lambda$CDM}    &  &  &  &  &  &  &  & \\[2mm]
     baseline+SDSS    & $0.3131\pm 0.0049$ & $67.41\pm 0.36$ & $-$ & $-$ & $-$ & $0.9653\pm 0.0036$ & $1.72^{+0.71}_{-1.3} $ & $-$ \\[2mm]
     baseline+DESI    & $0.3081\pm 0.0046$ & $67.77\pm 0.35$ & $-$ & $-$ & $-$ & $0.9673\pm 0.0035$ & $1.82^{+0.78}_{-1.3} $ &  $-$\\[2mm]
     SDSS+CMB+Union3    & $0.3133\pm 0.0050$ & $67.40\pm 0.37$ & $-$ & $-$ & $-$  & $0.9652\pm 0.0037$  & $1.72^{+0.71}_{-1.3} $ & $-$\\[2mm]
     DESI+CMB+Union3    & $0.3082\pm 0.0048$ & $67.77\pm 0.36$ & $-$ & $-$ & $-$ & $0.9673\pm 0.0036$ & $1.78^{+0.77}_{-1.3} $ & $-$\\[2mm]
     SDSS+CMB+DESY5    & $0.3152\pm 0.0049$ & $67.25\pm 0.36$ & $-$ & $-$ & $-$ & $0.9644\pm 0.0036$ & $1.69^{+0.69}_{-1.3} $ & $-$\\[2mm]
     DESI+CMB+DESY5    & $0.3101\pm 0.0047$ & $67.62\pm 0.35$ & $-$ & $-$ & $-$ & $0.9664\pm 0.0035$ & $1.76^{+0.73}_{-1.3} $  & $-$\\[2mm]
    \hline
    \textbf{$\Lambda$CDM $+ \Omega_K$}      &  &  &  &  &  &  &  & \\ [2mm]
     baseline+SDSS    & $0.3129\pm 0.0055$ & $67.44\pm 0.58$ & $0.1\pm 1.7$ & $-$ & $-$ & $0.9652\pm 0.0041$ & $1.72^{+0.67}_{-1.4} $ & $-$\\[2mm]
     baseline+DESI    & $0.3062^{+0.0045}_{-0.0050}$ & $68.27\pm 0.49$ & $2.2\pm 1.5$ & $-$ & $-$ & $0.9642^{+0.0044}_{-0.0038}$ & $1.73^{+0.60}_{-1.5} $ & $-$\\[2mm]
    \hline
    \textbf{$\Lambda_s$CDM} &  &  &  &  &  &  &  & \\[2mm]
     baseline+SDSS    & $0.3065\pm 0.0051$ & $68.35^{+0.41}_{-0.49}$  & $-$ & $-$ & $-$ & $0.9628\pm 0.0036$ & $1.66^{+0.66}_{-1.3} $ & $> 2.59$\\ [2mm]
     baseline+DESI    & $0.3027\pm 0.0047$ & $68.64\pm 0.40$ & $-$ & $-$ & $-$ & $0.9645\pm 0.0036$ & $1.67^{+0.67}_{-1.3} $ & $2.68^{+0.30}_{-0.12}$\\ [2mm]
    \hline
    \textbf{$\omega_0\omega_a$CDM}     &  &  &  &  &  &  &  & \\ [2mm]
     baseline+SDSS    & $0.3142\pm 0.0069$ & $67.34\pm 0.70$  & $-$ & $-0.882\pm 0.062$ & $-0.46^{+0.28}_{-0.23}$ & $0.9647\pm 0.0038$ & $1.71^{+0.68}_{-1.3} $ & $-$\\ [2mm]
     baseline+DESI    & $0.3080\pm 0.0068$ & $67.94\pm 0.72$ & $-$ & $-0.834\pm 0.062$ & $-0.71^{+0.28}_{-0.24}$ & $0.9655\pm 0.0037$ & $1.73^{+0.71}_{-1.3} $ & $-$\\ [2mm]
     SDSS+CMB+Union3    & $0.3275\pm0.094$ & $65.99\pm 0.92$ & $-$ & $-0.729\pm 0.094$ & $-0.87^{+0.35}_{-0.30}$ & $0.9645\pm 0.0038$ & $1.70^{+0.68}_{-1.3}$ & $-$\\ [2mm]
     DESI+CMB+Union3    & $0.3225\pm 0.0096$ & $66.44\pm 0.96$ & $-$ & $-0.66\pm 0.10$ & $-1.21^{+0.40}_{-0.34}$ & $0.9653\pm 0.0037$ & $1.70^{+0.66}_{-1.4} $ & $-$\\ [2mm]
     SDSS+CMB+DESY5    & $0.3212\pm 0.0066$  & $66.62\pm 0.65$  & $-$ & $-0.792\pm 0.064$ & $-0.71^{+0.29}_{-0.25}$  & $0.9646\pm 0.0038$ & $1.69^{+0.66}_{-1.4} $ & $-$\\ [2mm]
     DESI+CMB+DESY5    & $0.3157\pm 0.0065$ & $67.14\pm 0.67$ & $-$ & $-0.736\pm 0.069$  & $-1.00^{+0.32}_{-0.27}$ & $0.9653\pm 0.0037$ & $1.76^{+0.73}_{-1.3} $ & $-$\\ [2mm]
    \hline
    \textbf{wzlg-1}      &  &  &  &  &  &  &  &\\ [2mm]
     baseline+SDSS    & $0.3199\pm 0.0062$ & $66.57^{+0.59}_{-0.52}$  & $-$ & $< -0.959$ & $0.013^{+0.034}_{-0.056}$  & $0.9666\pm 0.0037$ & $1.78^{+0.73}_{-1.4} $ & $-$\\ [2mm]
     baseline+DESI    & $0.3146\pm 0.0058$ & $66.98^{+0.57}_{-0.49}$ & $-$ & $< -0.961$ & $0.005^{+0.031}_{-0.046}$ & $0.9684\pm 0.0036$ & $1.85^{+0.79}_{-1.3} $ & $-$\\ [2mm]
     SDSS+CMB+Union3    & $0.3248^{+0.0070}_{-0.0086}$ & $66.07^{+0.84}_{-0.66}$ & $-$ & $< -0.930$ & $-0.008^{+0.050}_{-0.068}  $ & $0.9669\pm 0.0037$ & $1.79^{+0.72}_{-1.3} $ & $-$\\ [2mm]
     DESI+CMB+Union3    & $0.3184^{+0.0065}_{-0.0080}$  & $66.58^{+0.80}_{-0.60}$ & $-$ & $< -0.939$  & $-0.012^{+0.045}_{-0.053}  $ & $0.9685\pm 0.0036$ & $1.86^{+0.78}_{-1.4} $ & $-$\\ [2mm]
     SDSS+CMB+DESY5    & $0.3252\pm 0.0064$ & $66.03\pm 0.59$ & $-$ & $-0.943^{+0.027}_{-0.035}  $  & $-0.009^{+0.040}_{-0.067}  $ & $0.9667\pm 0.0037$ & $1.79^{+0.71}_{-1.4} $ & $-$\\ [2mm]
     DESI+CMB+DESY5    & $0.3197\pm 0.0063$ & $66.43\pm 0.59$ & $-$ & $-0.943^{+0.026}_{-0.034}  $ & $-0.021^{+0.037}_{-0.054}  $ & $0.9686\pm 0.0036$ & $1.83^{+0.74}_{-1.3} $ & $-$\\ [2mm]
    \hline 
    \hline
    \end{tabular}}
    \caption{$68\%$ C.L. constraints on the cosmological parameters of interest for all the models studied, using different dataset combinations.}
    \label{tab:results}
\end{table*}

\section{Conclusions}
\label{sec:conclusions}

In this work, I have studied the impact of the DESI BAO data on key cosmological parameters, particularly the inflationary spectral index $n_s$, the amplitude of scalar perturbations $A_s$, and the matter density parameter $\omega_m$. The analyses show that the results are consistent with the flat $\Lambda$CDM model using the baseline+SDSS dataset, with all parameters remaining within a $1\sigma$ confidence level. The DESI data leads to a slight reduction in $\omega_m$ (by less than 2\%) and only modest shifts in $A_s$ and $n_s$, reflecting a subtle change in matter clustering.
These variations are consistent with late-time expansion effects, such as a higher expansion rate linked to dynamical dark energy or adjustments in the early Integrated Sachs-Wolfe amplitude to agree with CMB data.

We also explored models incorporating dynamical dark energy and free curvature alongside the $\Lambda$CDM model, finding that all models exhibit a consistent trend: a decrease in $\omega_m$ accompanied by slight increases in both $n_s$ and $H_0$. However, regardless of the model, the combination of BAO, CMB, and SNeIa data consistently favors lower values of $H_0$ (in the range of $66 - 68$ km/s/Mpc), which contrasts with local measurements.
Interestingly, the increase in $H_0$ observed in our study was previously reported in \cite{Pogosian:2024ykm}, where higher values of the product $r_d h$ were found, approximately $2\sigma$ away from the Planck best-fit $\Lambda$CDM value. Despite this, the DESI BAO data still agrees with the CMB acoustic scale, even without assuming a specific recombination model. This suggests that while there is limited flexibility in the treatment of recombination physics, minor deviations can be offset by adjustments in correlated parameters. These dynamics illustrate the compatibility of the various models considered, as well as the stability of inflationary parameters across them.

In conclusion, this study highlights the crucial role of DESI BAO data in refining cosmological parameter estimates, particularly in understanding the subtle shifts in matter density and inflationary parameters. Finally, it is essential to emphasize an important caveat of this work: the stability of the inflationary parameters in relation to late-time cosmological models. However, determining whether this trend will persist under different scenarios that incorporate further modifications is beyond the scope of the current study (see Refs. \cite{Giare:2024akf,Poulin:2023lkg} for recent reviews discussing possible implications for inflation resulting from considering new physics in light of the Hubble tension); thus, it would require analyses on a case-by-case basis (see Refs.\cite{Giare:2022rvg,Jiang:2022qlj,Ye:2022efx,Jiang:2022uyg} for more discussions on the theme). 

\begin{acknowledgments}
\noindent We acknowledge the use of high-performance computing services provided by the Observatório Nacional Data Center. S.S.C. acknowledges support from the Istituto Nazionale di Fisica Nucleare (INFN) through the Commissione Scientifica Nazionale 4 (CSN4) Iniziativa Specifica ``Quantum Fields in Gravity, Cosmology and Black Holes'' (FLAG) and from the Fondazione Cassa di Risparmio di Trento e Rovereto (CARITRO Foundation) through a Caritro Fellowship (project ``Inflation and dark sector physics in light of next-generation cosmological surveys'').
\end{acknowledgments}

\bibliography{infaftdesi}

\begin{thebibliography}{104}%
\makeatletter
\providecommand \@ifxundefined [1]{%
 \@ifx{#1\undefined}
}%
\providecommand \@ifnum [1]{%
 \ifnum #1\expandafter \@firstoftwo
 \else \expandafter \@secondoftwo
 \fi
}%
\providecommand \@ifx [1]{%
 \ifx #1\expandafter \@firstoftwo
 \else \expandafter \@secondoftwo
 \fi
}%
\providecommand \natexlab [1]{#1}%
\providecommand \enquote  [1]{``#1''}%
\providecommand \bibnamefont  [1]{#1}%
\providecommand \bibfnamefont [1]{#1}%
\providecommand \citenamefont [1]{#1}%
\providecommand \href@noop [0]{\@secondoftwo}%
\providecommand \href [0]{\begingroup \@sanitize@url \@href}%
\providecommand \@href[1]{\@@startlink{#1}\@@href}%
\providecommand \@@href[1]{\endgroup#1\@@endlink}%
\providecommand \@sanitize@url [0]{\catcode `\\12\catcode `\$12\catcode `\&12\catcode `\#12\catcode `\^12\catcode `\_12\catcode `\%12\relax}%
\providecommand \@@startlink[1]{}%
\providecommand \@@endlink[0]{}%
\providecommand \url  [0]{\begingroup\@sanitize@url \@url }%
\providecommand \@url [1]{\endgroup\@href {#1}{\urlprefix }}%
\providecommand \urlprefix  [0]{URL }%
\providecommand \Eprint [0]{\href }%
\providecommand \doibase [0]{http://dx.doi.org/}%
\providecommand \selectlanguage [0]{\@gobble}%
\providecommand \bibinfo  [0]{\@secondoftwo}%
\providecommand \bibfield  [0]{\@secondoftwo}%
\providecommand \translation [1]{[#1]}%
\providecommand \BibitemOpen [0]{}%
\providecommand \bibitemStop [0]{}%
\providecommand \bibitemNoStop [0]{.\EOS\space}%
\providecommand \EOS [0]{\spacefactor3000\relax}%
\providecommand \BibitemShut  [1]{\csname bibitem#1\endcsname}%
\let\auto@bib@innerbib\@empty
\bibitem [{\citenamefont {Riess}\ \emph {et~al.}(1998)\citenamefont {Riess} \emph {et~al.}}]{SupernovaSearchTeam:1998fmf}%
  \BibitemOpen
  \bibfield  {author} {\bibinfo {author} {\bibfnamefont {A.~G.}\ \bibnamefont {Riess}} \emph {et~al.} (\bibinfo {collaboration} {Supernova Search Team}),\ }\href {\doibase 10.1086/300499} {\bibfield  {journal} {\bibinfo  {journal} {Astron. J.}\ }\textbf {\bibinfo {volume} {116}},\ \bibinfo {pages} {1009} (\bibinfo {year} {1998})},\ \Eprint {http://arxiv.org/abs/astro-ph/9805201} {arXiv:astro-ph/9805201} \BibitemShut {NoStop}%
\bibitem [{\citenamefont {Perlmutter}\ \emph {et~al.}(1999)\citenamefont {Perlmutter} \emph {et~al.}}]{SupernovaCosmologyProject:1998vns}%
  \BibitemOpen
  \bibfield  {author} {\bibinfo {author} {\bibfnamefont {S.}~\bibnamefont {Perlmutter}} \emph {et~al.} (\bibinfo {collaboration} {Supernova Cosmology Project}),\ }\href {\doibase 10.1086/307221} {\bibfield  {journal} {\bibinfo  {journal} {Astrophys. J.}\ }\textbf {\bibinfo {volume} {517}},\ \bibinfo {pages} {565} (\bibinfo {year} {1999})},\ \Eprint {http://arxiv.org/abs/astro-ph/9812133} {arXiv:astro-ph/9812133} \BibitemShut {NoStop}%
\bibitem [{\citenamefont {Troxel}\ \emph {et~al.}(2018)\citenamefont {Troxel} \emph {et~al.}}]{DES:2017qwj}%
  \BibitemOpen
  \bibfield  {author} {\bibinfo {author} {\bibfnamefont {M.~A.}\ \bibnamefont {Troxel}} \emph {et~al.} (\bibinfo {collaboration} {DES}),\ }\href {\doibase 10.1103/PhysRevD.98.043528} {\bibfield  {journal} {\bibinfo  {journal} {Phys. Rev. D}\ }\textbf {\bibinfo {volume} {98}},\ \bibinfo {pages} {043528} (\bibinfo {year} {2018})},\ \Eprint {http://arxiv.org/abs/1708.01538} {arXiv:1708.01538 [astro-ph.CO]} \BibitemShut {NoStop}%
\bibitem [{\citenamefont {Aghanim}\ \emph {et~al.}(2020{\natexlab{a}})\citenamefont {Aghanim} \emph {et~al.}}]{Planck:2018vyg}%
  \BibitemOpen
  \bibfield  {author} {\bibinfo {author} {\bibfnamefont {N.}~\bibnamefont {Aghanim}} \emph {et~al.} (\bibinfo {collaboration} {Planck}),\ }\href {\doibase 10.1051/0004-6361/201833910} {\bibfield  {journal} {\bibinfo  {journal} {Astron. Astrophys.}\ }\textbf {\bibinfo {volume} {641}},\ \bibinfo {pages} {A6} (\bibinfo {year} {2020}{\natexlab{a}})},\ \bibinfo {note} {[Erratum: Astron.Astrophys. 652, C4 (2021)]},\ \Eprint {http://arxiv.org/abs/1807.06209} {arXiv:1807.06209 [astro-ph.CO]} \BibitemShut {NoStop}%
\bibitem [{\citenamefont {Bianchini}\ \emph {et~al.}(2020)\citenamefont {Bianchini} \emph {et~al.}}]{SPT:2019fqo}%
  \BibitemOpen
  \bibfield  {author} {\bibinfo {author} {\bibfnamefont {F.}~\bibnamefont {Bianchini}} \emph {et~al.} (\bibinfo {collaboration} {SPT}),\ }\href {\doibase 10.3847/1538-4357/ab6082} {\bibfield  {journal} {\bibinfo  {journal} {Astrophys. J.}\ }\textbf {\bibinfo {volume} {888}},\ \bibinfo {pages} {119} (\bibinfo {year} {2020})},\ \Eprint {http://arxiv.org/abs/1910.07157} {arXiv:1910.07157 [astro-ph.CO]} \BibitemShut {NoStop}%
\bibitem [{\citenamefont {Aiola}\ \emph {et~al.}(2020)\citenamefont {Aiola} \emph {et~al.}}]{ACT:2020gnv}%
  \BibitemOpen
  \bibfield  {author} {\bibinfo {author} {\bibfnamefont {S.}~\bibnamefont {Aiola}} \emph {et~al.} (\bibinfo {collaboration} {ACT}),\ }\href {\doibase 10.1088/1475-7516/2020/12/047} {\bibfield  {journal} {\bibinfo  {journal} {JCAP}\ }\textbf {\bibinfo {volume} {12}},\ \bibinfo {pages} {047} (\bibinfo {year} {2020})},\ \Eprint {http://arxiv.org/abs/2007.07288} {arXiv:2007.07288 [astro-ph.CO]} \BibitemShut {NoStop}%
\bibitem [{\citenamefont {Alam}\ \emph {et~al.}(2021)\citenamefont {Alam} \emph {et~al.}}]{eBOSS:2020yzd}%
  \BibitemOpen
  \bibfield  {author} {\bibinfo {author} {\bibfnamefont {S.}~\bibnamefont {Alam}} \emph {et~al.} (\bibinfo {collaboration} {eBOSS}),\ }\href {\doibase 10.1103/PhysRevD.103.083533} {\bibfield  {journal} {\bibinfo  {journal} {Phys. Rev. D}\ }\textbf {\bibinfo {volume} {103}},\ \bibinfo {pages} {083533} (\bibinfo {year} {2021})},\ \Eprint {http://arxiv.org/abs/2007.08991} {arXiv:2007.08991 [astro-ph.CO]} \BibitemShut {NoStop}%
\bibitem [{\citenamefont {Asgari}\ \emph {et~al.}(2021)\citenamefont {Asgari} \emph {et~al.}}]{KiDS:2020suj}%
  \BibitemOpen
  \bibfield  {author} {\bibinfo {author} {\bibfnamefont {M.}~\bibnamefont {Asgari}} \emph {et~al.} (\bibinfo {collaboration} {KiDS}),\ }\href {\doibase 10.1051/0004-6361/202039070} {\bibfield  {journal} {\bibinfo  {journal} {Astron. Astrophys.}\ }\textbf {\bibinfo {volume} {645}},\ \bibinfo {pages} {A104} (\bibinfo {year} {2021})},\ \Eprint {http://arxiv.org/abs/2007.15633} {arXiv:2007.15633 [astro-ph.CO]} \BibitemShut {NoStop}%
\bibitem [{\citenamefont {Mossa}\ \emph {et~al.}(2020)\citenamefont {Mossa} \emph {et~al.}}]{Mossa:2020gjc}%
  \BibitemOpen
  \bibfield  {author} {\bibinfo {author} {\bibfnamefont {V.}~\bibnamefont {Mossa}} \emph {et~al.},\ }\href {\doibase 10.1038/s41586-020-2878-4} {\bibfield  {journal} {\bibinfo  {journal} {Nature}\ }\textbf {\bibinfo {volume} {587}},\ \bibinfo {pages} {210} (\bibinfo {year} {2020})}\BibitemShut {NoStop}%
\bibitem [{\citenamefont {Brout}\ \emph {et~al.}(2022)\citenamefont {Brout} \emph {et~al.}}]{Brout:2022vxf}%
  \BibitemOpen
  \bibfield  {author} {\bibinfo {author} {\bibfnamefont {D.}~\bibnamefont {Brout}} \emph {et~al.},\ }\href {\doibase 10.3847/1538-4357/ac8e04} {\bibfield  {journal} {\bibinfo  {journal} {Astrophys. J.}\ }\textbf {\bibinfo {volume} {938}},\ \bibinfo {pages} {110} (\bibinfo {year} {2022})},\ \Eprint {http://arxiv.org/abs/2202.04077} {arXiv:2202.04077 [astro-ph.CO]} \BibitemShut {NoStop}%
\bibitem [{\citenamefont {Mukhanov}\ and\ \citenamefont {Chibisov}(1981)}]{Mukhanov:1981xt}%
  \BibitemOpen
  \bibfield  {author} {\bibinfo {author} {\bibfnamefont {V.~F.}\ \bibnamefont {Mukhanov}}\ and\ \bibinfo {author} {\bibfnamefont {G.~V.}\ \bibnamefont {Chibisov}},\ }\href@noop {} {\bibfield  {journal} {\bibinfo  {journal} {JETP Lett.}\ }\textbf {\bibinfo {volume} {33}},\ \bibinfo {pages} {532} (\bibinfo {year} {1981})}\BibitemShut {NoStop}%
\bibitem [{\citenamefont {Mukhanov}\ and\ \citenamefont {Chibisov}(1982)}]{Mukhanov:1982nu}%
  \BibitemOpen
  \bibfield  {author} {\bibinfo {author} {\bibfnamefont {V.~F.}\ \bibnamefont {Mukhanov}}\ and\ \bibinfo {author} {\bibfnamefont {G.~V.}\ \bibnamefont {Chibisov}},\ }\href@noop {} {\bibfield  {journal} {\bibinfo  {journal} {Sov. Phys. JETP}\ }\textbf {\bibinfo {volume} {56}},\ \bibinfo {pages} {258} (\bibinfo {year} {1982})}\BibitemShut {NoStop}%
\bibitem [{\citenamefont {Hawking}(1982)}]{Hawking:1982cz}%
  \BibitemOpen
  \bibfield  {author} {\bibinfo {author} {\bibfnamefont {S.~W.}\ \bibnamefont {Hawking}},\ }\href {\doibase 10.1016/0370-2693(82)90373-2} {\bibfield  {journal} {\bibinfo  {journal} {Phys. Lett. B}\ }\textbf {\bibinfo {volume} {115}},\ \bibinfo {pages} {295} (\bibinfo {year} {1982})}\BibitemShut {NoStop}%
\bibitem [{\citenamefont {Starobinsky}(1982)}]{Starobinsky:1982ee}%
  \BibitemOpen
  \bibfield  {author} {\bibinfo {author} {\bibfnamefont {A.~A.}\ \bibnamefont {Starobinsky}},\ }\href {\doibase 10.1016/0370-2693(82)90541-X} {\bibfield  {journal} {\bibinfo  {journal} {Phys. Lett. B}\ }\textbf {\bibinfo {volume} {117}},\ \bibinfo {pages} {175} (\bibinfo {year} {1982})}\BibitemShut {NoStop}%
\bibitem [{\citenamefont {Guth}\ and\ \citenamefont {Pi}(1982)}]{Guth:1982ec}%
  \BibitemOpen
  \bibfield  {author} {\bibinfo {author} {\bibfnamefont {A.~H.}\ \bibnamefont {Guth}}\ and\ \bibinfo {author} {\bibfnamefont {S.~Y.}\ \bibnamefont {Pi}},\ }\href {\doibase 10.1103/PhysRevLett.49.1110} {\bibfield  {journal} {\bibinfo  {journal} {Phys. Rev. Lett.}\ }\textbf {\bibinfo {volume} {49}},\ \bibinfo {pages} {1110} (\bibinfo {year} {1982})}\BibitemShut {NoStop}%
\bibitem [{\citenamefont {Bardeen}\ \emph {et~al.}(1983)\citenamefont {Bardeen}, \citenamefont {Steinhardt},\ and\ \citenamefont {Turner}}]{Bardeen:1983qw}%
  \BibitemOpen
  \bibfield  {author} {\bibinfo {author} {\bibfnamefont {J.~M.}\ \bibnamefont {Bardeen}}, \bibinfo {author} {\bibfnamefont {P.~J.}\ \bibnamefont {Steinhardt}}, \ and\ \bibinfo {author} {\bibfnamefont {M.~S.}\ \bibnamefont {Turner}},\ }\href {\doibase 10.1103/PhysRevD.28.679} {\bibfield  {journal} {\bibinfo  {journal} {Phys. Rev. D}\ }\textbf {\bibinfo {volume} {28}},\ \bibinfo {pages} {679} (\bibinfo {year} {1983})}\BibitemShut {NoStop}%
\bibitem [{\citenamefont {Martin}\ \emph {et~al.}(2014)\citenamefont {Martin}, \citenamefont {Ringeval},\ and\ \citenamefont {Vennin}}]{Martin:2013tda}%
  \BibitemOpen
  \bibfield  {author} {\bibinfo {author} {\bibfnamefont {J.}~\bibnamefont {Martin}}, \bibinfo {author} {\bibfnamefont {C.}~\bibnamefont {Ringeval}}, \ and\ \bibinfo {author} {\bibfnamefont {V.}~\bibnamefont {Vennin}},\ }\href {\doibase 10.1016/j.dark.2024.101653} {\bibfield  {journal} {\bibinfo  {journal} {Phys. Dark Univ.}\ }\textbf {\bibinfo {volume} {5-6}},\ \bibinfo {pages} {75} (\bibinfo {year} {2014})},\ \Eprint {http://arxiv.org/abs/1303.3787} {arXiv:1303.3787 [astro-ph.CO]} \BibitemShut {NoStop}%
\bibitem [{\citenamefont {Adame}\ \emph {et~al.}(2024)\citenamefont {Adame} \emph {et~al.}}]{DESI:2024mwx}%
  \BibitemOpen
  \bibfield  {author} {\bibinfo {author} {\bibfnamefont {A.~G.}\ \bibnamefont {Adame}} \emph {et~al.} (\bibinfo {collaboration} {DESI}),\ }\href@noop {} {\  (\bibinfo {year} {2024})},\ \Eprint {http://arxiv.org/abs/2404.03002} {arXiv:2404.03002 [astro-ph.CO]} \BibitemShut {NoStop}%
\bibitem [{\citenamefont {Perivolaropoulos}\ and\ \citenamefont {Skara}(2022)}]{Perivolaropoulos:2021jda}%
  \BibitemOpen
  \bibfield  {author} {\bibinfo {author} {\bibfnamefont {L.}~\bibnamefont {Perivolaropoulos}}\ and\ \bibinfo {author} {\bibfnamefont {F.}~\bibnamefont {Skara}},\ }\href {\doibase 10.1016/j.newar.2022.101659} {\bibfield  {journal} {\bibinfo  {journal} {New Astron. Rev.}\ }\textbf {\bibinfo {volume} {95}},\ \bibinfo {pages} {101659} (\bibinfo {year} {2022})},\ \Eprint {http://arxiv.org/abs/2105.05208} {arXiv:2105.05208 [astro-ph.CO]} \BibitemShut {NoStop}%
\bibitem [{\citenamefont {Abdalla}\ \emph {et~al.}(2022)\citenamefont {Abdalla} \emph {et~al.}}]{Abdalla:2022yfr}%
  \BibitemOpen
  \bibfield  {author} {\bibinfo {author} {\bibfnamefont {E.}~\bibnamefont {Abdalla}} \emph {et~al.},\ }\href {\doibase 10.1016/j.jheap.2022.04.002} {\bibfield  {journal} {\bibinfo  {journal} {JHEAp}\ }\textbf {\bibinfo {volume} {34}},\ \bibinfo {pages} {49} (\bibinfo {year} {2022})},\ \Eprint {http://arxiv.org/abs/2203.06142} {arXiv:2203.06142 [astro-ph.CO]} \BibitemShut {NoStop}%
\bibitem [{\citenamefont {Akarsu}\ \emph {et~al.}(2024)\citenamefont {Akarsu}, \citenamefont {Colg\'ain}, \citenamefont {Sen},\ and\ \citenamefont {Sheikh-Jabbari}}]{Akarsu:2024qiq}%
  \BibitemOpen
  \bibfield  {author} {\bibinfo {author} {\bibfnamefont {O.}~\bibnamefont {Akarsu}}, \bibinfo {author} {\bibfnamefont {E.~O.}\ \bibnamefont {Colg\'ain}}, \bibinfo {author} {\bibfnamefont {A.~A.}\ \bibnamefont {Sen}}, \ and\ \bibinfo {author} {\bibfnamefont {M.~M.}\ \bibnamefont {Sheikh-Jabbari}},\ }\href {\doibase 10.3390/universe10080305} {\bibfield  {journal} {\bibinfo  {journal} {Universe}\ }\textbf {\bibinfo {volume} {10}},\ \bibinfo {pages} {305} (\bibinfo {year} {2024})},\ \Eprint {http://arxiv.org/abs/2402.04767} {arXiv:2402.04767 [astro-ph.CO]} \BibitemShut {NoStop}%
\bibitem [{\citenamefont {Bernal}\ \emph {et~al.}(2016)\citenamefont {Bernal}, \citenamefont {Verde},\ and\ \citenamefont {Riess}}]{Bernal:2016gxb}%
  \BibitemOpen
  \bibfield  {author} {\bibinfo {author} {\bibfnamefont {J.~L.}\ \bibnamefont {Bernal}}, \bibinfo {author} {\bibfnamefont {L.}~\bibnamefont {Verde}}, \ and\ \bibinfo {author} {\bibfnamefont {A.~G.}\ \bibnamefont {Riess}},\ }\href {\doibase 10.1088/1475-7516/2016/10/019} {\bibfield  {journal} {\bibinfo  {journal} {JCAP}\ }\textbf {\bibinfo {volume} {10}},\ \bibinfo {pages} {019} (\bibinfo {year} {2016})},\ \Eprint {http://arxiv.org/abs/1607.05617} {arXiv:1607.05617 [astro-ph.CO]} \BibitemShut {NoStop}%
\bibitem [{\citenamefont {Addison}\ \emph {et~al.}(2018)\citenamefont {Addison}, \citenamefont {Watts}, \citenamefont {Bennett}, \citenamefont {Halpern}, \citenamefont {Hinshaw},\ and\ \citenamefont {Weiland}}]{Addison:2017fdm}%
  \BibitemOpen
  \bibfield  {author} {\bibinfo {author} {\bibfnamefont {G.~E.}\ \bibnamefont {Addison}}, \bibinfo {author} {\bibfnamefont {D.~J.}\ \bibnamefont {Watts}}, \bibinfo {author} {\bibfnamefont {C.~L.}\ \bibnamefont {Bennett}}, \bibinfo {author} {\bibfnamefont {M.}~\bibnamefont {Halpern}}, \bibinfo {author} {\bibfnamefont {G.}~\bibnamefont {Hinshaw}}, \ and\ \bibinfo {author} {\bibfnamefont {J.~L.}\ \bibnamefont {Weiland}},\ }\href {\doibase 10.3847/1538-4357/aaa1ed} {\bibfield  {journal} {\bibinfo  {journal} {Astrophys. J.}\ }\textbf {\bibinfo {volume} {853}},\ \bibinfo {pages} {119} (\bibinfo {year} {2018})},\ \Eprint {http://arxiv.org/abs/1707.06547} {arXiv:1707.06547 [astro-ph.CO]} \BibitemShut {NoStop}%
\bibitem [{\citenamefont {Lemos}\ \emph {et~al.}(2019)\citenamefont {Lemos}, \citenamefont {Lee}, \citenamefont {Efstathiou},\ and\ \citenamefont {Gratton}}]{Lemos:2018smw}%
  \BibitemOpen
  \bibfield  {author} {\bibinfo {author} {\bibfnamefont {P.}~\bibnamefont {Lemos}}, \bibinfo {author} {\bibfnamefont {E.}~\bibnamefont {Lee}}, \bibinfo {author} {\bibfnamefont {G.}~\bibnamefont {Efstathiou}}, \ and\ \bibinfo {author} {\bibfnamefont {S.}~\bibnamefont {Gratton}},\ }\href {\doibase 10.1093/mnras/sty3082} {\bibfield  {journal} {\bibinfo  {journal} {Mon. Not. Roy. Astron. Soc.}\ }\textbf {\bibinfo {volume} {483}},\ \bibinfo {pages} {4803} (\bibinfo {year} {2019})},\ \Eprint {http://arxiv.org/abs/1806.06781} {arXiv:1806.06781 [astro-ph.CO]} \BibitemShut {NoStop}%
\bibitem [{\citenamefont {Aylor}\ \emph {et~al.}(2019)\citenamefont {Aylor}, \citenamefont {Joy}, \citenamefont {Knox}, \citenamefont {Millea}, \citenamefont {Raghunathan},\ and\ \citenamefont {Wu}}]{Aylor:2018drw}%
  \BibitemOpen
  \bibfield  {author} {\bibinfo {author} {\bibfnamefont {K.}~\bibnamefont {Aylor}}, \bibinfo {author} {\bibfnamefont {M.}~\bibnamefont {Joy}}, \bibinfo {author} {\bibfnamefont {L.}~\bibnamefont {Knox}}, \bibinfo {author} {\bibfnamefont {M.}~\bibnamefont {Millea}}, \bibinfo {author} {\bibfnamefont {S.}~\bibnamefont {Raghunathan}}, \ and\ \bibinfo {author} {\bibfnamefont {W.~L.~K.}\ \bibnamefont {Wu}},\ }\href {\doibase 10.3847/1538-4357/ab0898} {\bibfield  {journal} {\bibinfo  {journal} {Astrophys. J.}\ }\textbf {\bibinfo {volume} {874}},\ \bibinfo {pages} {4} (\bibinfo {year} {2019})},\ \Eprint {http://arxiv.org/abs/1811.00537} {arXiv:1811.00537 [astro-ph.CO]} \BibitemShut {NoStop}%
\bibitem [{\citenamefont {Knox}\ and\ \citenamefont {Millea}(2020)}]{Knox:2019rjx}%
  \BibitemOpen
  \bibfield  {author} {\bibinfo {author} {\bibfnamefont {L.}~\bibnamefont {Knox}}\ and\ \bibinfo {author} {\bibfnamefont {M.}~\bibnamefont {Millea}},\ }\href {\doibase 10.1103/PhysRevD.101.043533} {\bibfield  {journal} {\bibinfo  {journal} {Phys. Rev. D}\ }\textbf {\bibinfo {volume} {101}},\ \bibinfo {pages} {043533} (\bibinfo {year} {2020})},\ \Eprint {http://arxiv.org/abs/1908.03663} {arXiv:1908.03663 [astro-ph.CO]} \BibitemShut {NoStop}%
\bibitem [{\citenamefont {Arendse}\ \emph {et~al.}(2020)\citenamefont {Arendse} \emph {et~al.}}]{Arendse:2019hev}%
  \BibitemOpen
  \bibfield  {author} {\bibinfo {author} {\bibfnamefont {N.}~\bibnamefont {Arendse}} \emph {et~al.},\ }\href {\doibase 10.1051/0004-6361/201936720} {\bibfield  {journal} {\bibinfo  {journal} {Astron. Astrophys.}\ }\textbf {\bibinfo {volume} {639}},\ \bibinfo {pages} {A57} (\bibinfo {year} {2020})},\ \Eprint {http://arxiv.org/abs/1909.07986} {arXiv:1909.07986 [astro-ph.CO]} \BibitemShut {NoStop}%
\bibitem [{\citenamefont {Efstathiou}(2021)}]{Efstathiou:2021ocp}%
  \BibitemOpen
  \bibfield  {author} {\bibinfo {author} {\bibfnamefont {G.}~\bibnamefont {Efstathiou}},\ }\href {\doibase 10.1093/mnras/stab1588} {\bibfield  {journal} {\bibinfo  {journal} {Mon. Not. Roy. Astron. Soc.}\ }\textbf {\bibinfo {volume} {505}},\ \bibinfo {pages} {3866} (\bibinfo {year} {2021})},\ \Eprint {http://arxiv.org/abs/2103.08723} {arXiv:2103.08723 [astro-ph.CO]} \BibitemShut {NoStop}%
\bibitem [{\citenamefont {Cai}\ \emph {et~al.}(2022)\citenamefont {Cai}, \citenamefont {Guo}, \citenamefont {Wang}, \citenamefont {Yu},\ and\ \citenamefont {Zhou}}]{Cai:2021weh}%
  \BibitemOpen
  \bibfield  {author} {\bibinfo {author} {\bibfnamefont {R.-G.}\ \bibnamefont {Cai}}, \bibinfo {author} {\bibfnamefont {Z.-K.}\ \bibnamefont {Guo}}, \bibinfo {author} {\bibfnamefont {S.-J.}\ \bibnamefont {Wang}}, \bibinfo {author} {\bibfnamefont {W.-W.}\ \bibnamefont {Yu}}, \ and\ \bibinfo {author} {\bibfnamefont {Y.}~\bibnamefont {Zhou}},\ }\href {\doibase 10.1103/PhysRevD.105.L021301} {\bibfield  {journal} {\bibinfo  {journal} {Phys. Rev. D}\ }\textbf {\bibinfo {volume} {105}},\ \bibinfo {pages} {L021301} (\bibinfo {year} {2022})},\ \Eprint {http://arxiv.org/abs/2107.13286} {arXiv:2107.13286 [astro-ph.CO]} \BibitemShut {NoStop}%
\bibitem [{\citenamefont {Keeley}\ and\ \citenamefont {Shafieloo}(2023)}]{Keeley:2022ojz}%
  \BibitemOpen
  \bibfield  {author} {\bibinfo {author} {\bibfnamefont {R.~E.}\ \bibnamefont {Keeley}}\ and\ \bibinfo {author} {\bibfnamefont {A.}~\bibnamefont {Shafieloo}},\ }\href {\doibase 10.1103/PhysRevLett.131.111002} {\bibfield  {journal} {\bibinfo  {journal} {Phys. Rev. Lett.}\ }\textbf {\bibinfo {volume} {131}},\ \bibinfo {pages} {111002} (\bibinfo {year} {2023})},\ \Eprint {http://arxiv.org/abs/2206.08440} {arXiv:2206.08440 [astro-ph.CO]} \BibitemShut {NoStop}%
\bibitem [{\citenamefont {Vagnozzi}(2023)}]{Vagnozzi:2023nrq}%
  \BibitemOpen
  \bibfield  {author} {\bibinfo {author} {\bibfnamefont {S.}~\bibnamefont {Vagnozzi}},\ }\href {\doibase 10.3390/universe9090393} {\bibfield  {journal} {\bibinfo  {journal} {Universe}\ }\textbf {\bibinfo {volume} {9}},\ \bibinfo {pages} {393} (\bibinfo {year} {2023})},\ \Eprint {http://arxiv.org/abs/2308.16628} {arXiv:2308.16628 [astro-ph.CO]} \BibitemShut {NoStop}%
\bibitem [{\citenamefont {Lin}\ \emph {et~al.}(2020)\citenamefont {Lin}, \citenamefont {Hu},\ and\ \citenamefont {Raveri}}]{Lin:2020jcb}%
  \BibitemOpen
  \bibfield  {author} {\bibinfo {author} {\bibfnamefont {M.-X.}\ \bibnamefont {Lin}}, \bibinfo {author} {\bibfnamefont {W.}~\bibnamefont {Hu}}, \ and\ \bibinfo {author} {\bibfnamefont {M.}~\bibnamefont {Raveri}},\ }\href {\doibase 10.1103/PhysRevD.102.123523} {\bibfield  {journal} {\bibinfo  {journal} {Phys. Rev. D}\ }\textbf {\bibinfo {volume} {102}},\ \bibinfo {pages} {123523} (\bibinfo {year} {2020})},\ \Eprint {http://arxiv.org/abs/2009.08974} {arXiv:2009.08974 [astro-ph.CO]} \BibitemShut {NoStop}%
\bibitem [{\citenamefont {McDonough}\ \emph {et~al.}(2024)\citenamefont {McDonough}, \citenamefont {Hill}, \citenamefont {Ivanov}, \citenamefont {La~Posta},\ and\ \citenamefont {Toomey}}]{McDonough:2023qcu}%
  \BibitemOpen
  \bibfield  {author} {\bibinfo {author} {\bibfnamefont {E.}~\bibnamefont {McDonough}}, \bibinfo {author} {\bibfnamefont {J.~C.}\ \bibnamefont {Hill}}, \bibinfo {author} {\bibfnamefont {M.~M.}\ \bibnamefont {Ivanov}}, \bibinfo {author} {\bibfnamefont {A.}~\bibnamefont {La~Posta}}, \ and\ \bibinfo {author} {\bibfnamefont {M.~W.}\ \bibnamefont {Toomey}},\ }\href {\doibase 10.1142/S0218271824300039} {\bibfield  {journal} {\bibinfo  {journal} {Int. J. Mod. Phys. D}\ }\textbf {\bibinfo {volume} {33}},\ \bibinfo {pages} {2430003} (\bibinfo {year} {2024})},\ \Eprint {http://arxiv.org/abs/2310.19899} {arXiv:2310.19899 [astro-ph.CO]} \BibitemShut {NoStop}%
\bibitem [{\citenamefont {Simon}(2024)}]{Simon:2023hlp}%
  \BibitemOpen
  \bibfield  {author} {\bibinfo {author} {\bibfnamefont {T.}~\bibnamefont {Simon}},\ }\href {\doibase 10.1103/PhysRevD.110.023528} {\bibfield  {journal} {\bibinfo  {journal} {Phys. Rev. D}\ }\textbf {\bibinfo {volume} {110}},\ \bibinfo {pages} {023528} (\bibinfo {year} {2024})},\ \Eprint {http://arxiv.org/abs/2310.16800} {arXiv:2310.16800 [astro-ph.CO]} \BibitemShut {NoStop}%
\bibitem [{\citenamefont {Pedrotti}\ \emph {et~al.}(2024)\citenamefont {Pedrotti}, \citenamefont {Jiang}, \citenamefont {Escamilla}, \citenamefont {da~Costa},\ and\ \citenamefont {Vagnozzi}}]{Pedrotti:2024kpn}%
  \BibitemOpen
  \bibfield  {author} {\bibinfo {author} {\bibfnamefont {D.}~\bibnamefont {Pedrotti}}, \bibinfo {author} {\bibfnamefont {J.-Q.}\ \bibnamefont {Jiang}}, \bibinfo {author} {\bibfnamefont {L.~A.}\ \bibnamefont {Escamilla}}, \bibinfo {author} {\bibfnamefont {S.~S.}\ \bibnamefont {da~Costa}}, \ and\ \bibinfo {author} {\bibfnamefont {S.}~\bibnamefont {Vagnozzi}},\ }\href@noop {} {\  (\bibinfo {year} {2024})},\ \Eprint {http://arxiv.org/abs/2408.04530} {arXiv:2408.04530 [astro-ph.CO]} \BibitemShut {NoStop}%
\bibitem [{\citenamefont {Alam}\ \emph {et~al.}(2017)\citenamefont {Alam} \emph {et~al.}}]{BOSS:2016wmc}%
  \BibitemOpen
  \bibfield  {author} {\bibinfo {author} {\bibfnamefont {S.}~\bibnamefont {Alam}} \emph {et~al.} (\bibinfo {collaboration} {BOSS}),\ }\href {\doibase 10.1093/mnras/stx721} {\bibfield  {journal} {\bibinfo  {journal} {Mon. Not. Roy. Astron. Soc.}\ }\textbf {\bibinfo {volume} {470}},\ \bibinfo {pages} {2617} (\bibinfo {year} {2017})},\ \Eprint {http://arxiv.org/abs/1607.03155} {arXiv:1607.03155 [astro-ph.CO]} \BibitemShut {NoStop}%
\bibitem [{\citenamefont {Jiang}\ \emph {et~al.}(2024{\natexlab{a}})\citenamefont {Jiang}, \citenamefont {Ye},\ and\ \citenamefont {Piao}}]{Jiang:2023bsz}%
  \BibitemOpen
  \bibfield  {author} {\bibinfo {author} {\bibfnamefont {J.-Q.}\ \bibnamefont {Jiang}}, \bibinfo {author} {\bibfnamefont {G.}~\bibnamefont {Ye}}, \ and\ \bibinfo {author} {\bibfnamefont {Y.-S.}\ \bibnamefont {Piao}},\ }\href {\doibase 10.1016/j.physletb.2024.138588} {\bibfield  {journal} {\bibinfo  {journal} {Phys. Lett. B}\ }\textbf {\bibinfo {volume} {851}},\ \bibinfo {pages} {138588} (\bibinfo {year} {2024}{\natexlab{a}})},\ \Eprint {http://arxiv.org/abs/2303.12345} {arXiv:2303.12345 [astro-ph.CO]} \BibitemShut {NoStop}%
\bibitem [{\citenamefont {Jiang}(2024)}]{Jiang:2024nha}%
  \BibitemOpen
  \bibfield  {author} {\bibinfo {author} {\bibfnamefont {J.-Q.}\ \bibnamefont {Jiang}},\ }\href@noop {} {\  (\bibinfo {year} {2024})},\ \Eprint {http://arxiv.org/abs/2410.10559} {arXiv:2410.10559 [astro-ph.CO]} \BibitemShut {NoStop}%
\bibitem [{\citenamefont {Wang}\ and\ \citenamefont {Piao}(2024)}]{Wang:2024dka}%
  \BibitemOpen
  \bibfield  {author} {\bibinfo {author} {\bibfnamefont {H.}~\bibnamefont {Wang}}\ and\ \bibinfo {author} {\bibfnamefont {Y.-S.}\ \bibnamefont {Piao}},\ }\href@noop {} {\  (\bibinfo {year} {2024})},\ \Eprint {http://arxiv.org/abs/2404.18579} {arXiv:2404.18579 [astro-ph.CO]} \BibitemShut {NoStop}%
\bibitem [{\citenamefont {Lynch}\ \emph {et~al.}(2024)\citenamefont {Lynch}, \citenamefont {Knox},\ and\ \citenamefont {Chluba}}]{Lynch:2024hzh}%
  \BibitemOpen
  \bibfield  {author} {\bibinfo {author} {\bibfnamefont {G.~P.}\ \bibnamefont {Lynch}}, \bibinfo {author} {\bibfnamefont {L.}~\bibnamefont {Knox}}, \ and\ \bibinfo {author} {\bibfnamefont {J.}~\bibnamefont {Chluba}},\ }\href {\doibase 10.1103/PhysRevD.110.083538} {\bibfield  {journal} {\bibinfo  {journal} {Phys. Rev. D}\ }\textbf {\bibinfo {volume} {110}},\ \bibinfo {pages} {083538} (\bibinfo {year} {2024})},\ \Eprint {http://arxiv.org/abs/2406.10202} {arXiv:2406.10202 [astro-ph.CO]} \BibitemShut {NoStop}%
\bibitem [{\citenamefont {Di~Valentino}\ \emph {et~al.}(2016)\citenamefont {Di~Valentino}, \citenamefont {Melchiorri},\ and\ \citenamefont {Silk}}]{DiValentino:2016hlg}%
  \BibitemOpen
  \bibfield  {author} {\bibinfo {author} {\bibfnamefont {E.}~\bibnamefont {Di~Valentino}}, \bibinfo {author} {\bibfnamefont {A.}~\bibnamefont {Melchiorri}}, \ and\ \bibinfo {author} {\bibfnamefont {J.}~\bibnamefont {Silk}},\ }\href {\doibase 10.1016/j.physletb.2016.08.043} {\bibfield  {journal} {\bibinfo  {journal} {Phys. Lett. B}\ }\textbf {\bibinfo {volume} {761}},\ \bibinfo {pages} {242} (\bibinfo {year} {2016})},\ \Eprint {http://arxiv.org/abs/1606.00634} {arXiv:1606.00634 [astro-ph.CO]} \BibitemShut {NoStop}%
\bibitem [{\citenamefont {Vagnozzi}(2020)}]{Vagnozzi:2019ezj}%
  \BibitemOpen
  \bibfield  {author} {\bibinfo {author} {\bibfnamefont {S.}~\bibnamefont {Vagnozzi}},\ }\href {\doibase 10.1103/PhysRevD.102.023518} {\bibfield  {journal} {\bibinfo  {journal} {Phys. Rev. D}\ }\textbf {\bibinfo {volume} {102}},\ \bibinfo {pages} {023518} (\bibinfo {year} {2020})},\ \Eprint {http://arxiv.org/abs/1907.07569} {arXiv:1907.07569 [astro-ph.CO]} \BibitemShut {NoStop}%
\bibitem [{\citenamefont {Tada}\ and\ \citenamefont {Terada}(2024)}]{Tada:2024znt}%
  \BibitemOpen
  \bibfield  {author} {\bibinfo {author} {\bibfnamefont {Y.}~\bibnamefont {Tada}}\ and\ \bibinfo {author} {\bibfnamefont {T.}~\bibnamefont {Terada}},\ }\href {\doibase 10.1103/PhysRevD.109.L121305} {\bibfield  {journal} {\bibinfo  {journal} {Phys. Rev. D}\ }\textbf {\bibinfo {volume} {109}},\ \bibinfo {pages} {L121305} (\bibinfo {year} {2024})},\ \Eprint {http://arxiv.org/abs/2404.05722} {arXiv:2404.05722 [astro-ph.CO]} \BibitemShut {NoStop}%
\bibitem [{\citenamefont {Ramadan}\ \emph {et~al.}(2024)\citenamefont {Ramadan}, \citenamefont {Sakstein},\ and\ \citenamefont {Rubin}}]{Ramadan:2024kmn}%
  \BibitemOpen
  \bibfield  {author} {\bibinfo {author} {\bibfnamefont {O.~F.}\ \bibnamefont {Ramadan}}, \bibinfo {author} {\bibfnamefont {J.}~\bibnamefont {Sakstein}}, \ and\ \bibinfo {author} {\bibfnamefont {D.}~\bibnamefont {Rubin}},\ }\href {\doibase 10.1103/PhysRevD.110.L041303} {\bibfield  {journal} {\bibinfo  {journal} {Phys. Rev. D}\ }\textbf {\bibinfo {volume} {110}},\ \bibinfo {pages} {L041303} (\bibinfo {year} {2024})},\ \Eprint {http://arxiv.org/abs/2405.18747} {arXiv:2405.18747 [astro-ph.CO]} \BibitemShut {NoStop}%
\bibitem [{\citenamefont {Bhattacharya}\ \emph {et~al.}(2024)\citenamefont {Bhattacharya}, \citenamefont {Borghetto}, \citenamefont {Malhotra}, \citenamefont {Parameswaran}, \citenamefont {Tasinato},\ and\ \citenamefont {Zavala}}]{Bhattacharya:2024hep}%
  \BibitemOpen
  \bibfield  {author} {\bibinfo {author} {\bibfnamefont {S.}~\bibnamefont {Bhattacharya}}, \bibinfo {author} {\bibfnamefont {G.}~\bibnamefont {Borghetto}}, \bibinfo {author} {\bibfnamefont {A.}~\bibnamefont {Malhotra}}, \bibinfo {author} {\bibfnamefont {S.}~\bibnamefont {Parameswaran}}, \bibinfo {author} {\bibfnamefont {G.}~\bibnamefont {Tasinato}}, \ and\ \bibinfo {author} {\bibfnamefont {I.}~\bibnamefont {Zavala}},\ }\href {\doibase 10.1088/1475-7516/2024/09/073} {\bibfield  {journal} {\bibinfo  {journal} {JCAP}\ }\textbf {\bibinfo {volume} {09}},\ \bibinfo {pages} {073} (\bibinfo {year} {2024})},\ \Eprint {http://arxiv.org/abs/2405.17396} {arXiv:2405.17396 [astro-ph.CO]} \BibitemShut {NoStop}%
\bibitem [{\citenamefont {Giar\`e}\ \emph {et~al.}(2024)\citenamefont {Giar\`e}, \citenamefont {Sabogal}, \citenamefont {Nunes},\ and\ \citenamefont {Di~Valentino}}]{Giare:2024smz}%
  \BibitemOpen
  \bibfield  {author} {\bibinfo {author} {\bibfnamefont {W.}~\bibnamefont {Giar\`e}}, \bibinfo {author} {\bibfnamefont {M.~A.}\ \bibnamefont {Sabogal}}, \bibinfo {author} {\bibfnamefont {R.~C.}\ \bibnamefont {Nunes}}, \ and\ \bibinfo {author} {\bibfnamefont {E.}~\bibnamefont {Di~Valentino}},\ }\href@noop {} {\  (\bibinfo {year} {2024})},\ \Eprint {http://arxiv.org/abs/2404.15232} {arXiv:2404.15232 [astro-ph.CO]} \BibitemShut {NoStop}%
\bibitem [{\citenamefont {Giar\`e}(2024{\natexlab{a}})}]{Giare:2024ocw}%
  \BibitemOpen
  \bibfield  {author} {\bibinfo {author} {\bibfnamefont {W.}~\bibnamefont {Giar\`e}},\ }\href@noop {} {\  (\bibinfo {year} {2024}{\natexlab{a}})},\ \Eprint {http://arxiv.org/abs/2409.17074} {arXiv:2409.17074 [astro-ph.CO]} \BibitemShut {NoStop}%
\bibitem [{\citenamefont {Jiang}\ \emph {et~al.}(2024{\natexlab{b}})\citenamefont {Jiang}, \citenamefont {Giar\`e}, \citenamefont {Gariazzo}, \citenamefont {Dainotti}, \citenamefont {Di~Valentino}, \citenamefont {Mena}, \citenamefont {Pedrotti}, \citenamefont {da~Costa},\ and\ \citenamefont {Vagnozzi}}]{Jiang:2024viw}%
  \BibitemOpen
  \bibfield  {author} {\bibinfo {author} {\bibfnamefont {J.-Q.}\ \bibnamefont {Jiang}}, \bibinfo {author} {\bibfnamefont {W.}~\bibnamefont {Giar\`e}}, \bibinfo {author} {\bibfnamefont {S.}~\bibnamefont {Gariazzo}}, \bibinfo {author} {\bibfnamefont {M.~G.}\ \bibnamefont {Dainotti}}, \bibinfo {author} {\bibfnamefont {E.}~\bibnamefont {Di~Valentino}}, \bibinfo {author} {\bibfnamefont {O.}~\bibnamefont {Mena}}, \bibinfo {author} {\bibfnamefont {D.}~\bibnamefont {Pedrotti}}, \bibinfo {author} {\bibfnamefont {S.~S.}\ \bibnamefont {da~Costa}}, \ and\ \bibinfo {author} {\bibfnamefont {S.}~\bibnamefont {Vagnozzi}},\ }\href@noop {} {\  (\bibinfo {year} {2024}{\natexlab{b}})},\ \Eprint {http://arxiv.org/abs/2407.18047} {arXiv:2407.18047 [astro-ph.CO]} \BibitemShut {NoStop}%
\bibitem [{\citenamefont {Naredo-Tuero}\ \emph {et~al.}(2024)\citenamefont {Naredo-Tuero}, \citenamefont {Escudero}, \citenamefont {Fern\'andez-Mart\'\i{}nez}, \citenamefont {Marcano},\ and\ \citenamefont {Poulin}}]{Naredo-Tuero:2024sgf}%
  \BibitemOpen
  \bibfield  {author} {\bibinfo {author} {\bibfnamefont {D.}~\bibnamefont {Naredo-Tuero}}, \bibinfo {author} {\bibfnamefont {M.}~\bibnamefont {Escudero}}, \bibinfo {author} {\bibfnamefont {E.}~\bibnamefont {Fern\'andez-Mart\'\i{}nez}}, \bibinfo {author} {\bibfnamefont {X.}~\bibnamefont {Marcano}}, \ and\ \bibinfo {author} {\bibfnamefont {V.}~\bibnamefont {Poulin}},\ }\href@noop {} {\  (\bibinfo {year} {2024})},\ \Eprint {http://arxiv.org/abs/2407.13831} {arXiv:2407.13831 [astro-ph.CO]} \BibitemShut {NoStop}%
\bibitem [{\citenamefont {Chudaykin}\ and\ \citenamefont {Kunz}(2024)}]{Chudaykin:2024gol}%
  \BibitemOpen
  \bibfield  {author} {\bibinfo {author} {\bibfnamefont {A.}~\bibnamefont {Chudaykin}}\ and\ \bibinfo {author} {\bibfnamefont {M.}~\bibnamefont {Kunz}},\ }\href@noop {} {\  (\bibinfo {year} {2024})},\ \Eprint {http://arxiv.org/abs/2407.02558} {arXiv:2407.02558 [astro-ph.CO]} \BibitemShut {NoStop}%
\bibitem [{\citenamefont {Beutler}\ \emph {et~al.}(2011)\citenamefont {Beutler}, \citenamefont {Blake}, \citenamefont {Colless}, \citenamefont {Jones}, \citenamefont {Staveley-Smith}, \citenamefont {Campbell}, \citenamefont {Parker}, \citenamefont {Saunders},\ and\ \citenamefont {Watson}}]{Beutler:2011hx}%
  \BibitemOpen
  \bibfield  {author} {\bibinfo {author} {\bibfnamefont {F.}~\bibnamefont {Beutler}}, \bibinfo {author} {\bibfnamefont {C.}~\bibnamefont {Blake}}, \bibinfo {author} {\bibfnamefont {M.}~\bibnamefont {Colless}}, \bibinfo {author} {\bibfnamefont {D.~H.}\ \bibnamefont {Jones}}, \bibinfo {author} {\bibfnamefont {L.}~\bibnamefont {Staveley-Smith}}, \bibinfo {author} {\bibfnamefont {L.}~\bibnamefont {Campbell}}, \bibinfo {author} {\bibfnamefont {Q.}~\bibnamefont {Parker}}, \bibinfo {author} {\bibfnamefont {W.}~\bibnamefont {Saunders}}, \ and\ \bibinfo {author} {\bibfnamefont {F.}~\bibnamefont {Watson}},\ }\href {\doibase 10.1111/j.1365-2966.2011.19250.x} {\bibfield  {journal} {\bibinfo  {journal} {Mon. Not. Roy. Astron. Soc.}\ }\textbf {\bibinfo {volume} {416}},\ \bibinfo {pages} {3017} (\bibinfo {year} {2011})},\ \Eprint {http://arxiv.org/abs/1106.3366} {arXiv:1106.3366 [astro-ph.CO]} \BibitemShut {NoStop}%
\bibitem [{\citenamefont {Ross}\ \emph {et~al.}(2015)\citenamefont {Ross}, \citenamefont {Samushia}, \citenamefont {Howlett}, \citenamefont {Percival}, \citenamefont {Burden},\ and\ \citenamefont {Manera}}]{Ross:2014qpa}%
  \BibitemOpen
  \bibfield  {author} {\bibinfo {author} {\bibfnamefont {A.~J.}\ \bibnamefont {Ross}}, \bibinfo {author} {\bibfnamefont {L.}~\bibnamefont {Samushia}}, \bibinfo {author} {\bibfnamefont {C.}~\bibnamefont {Howlett}}, \bibinfo {author} {\bibfnamefont {W.~J.}\ \bibnamefont {Percival}}, \bibinfo {author} {\bibfnamefont {A.}~\bibnamefont {Burden}}, \ and\ \bibinfo {author} {\bibfnamefont {M.}~\bibnamefont {Manera}},\ }\href {\doibase 10.1093/mnras/stv154} {\bibfield  {journal} {\bibinfo  {journal} {Mon. Not. Roy. Astron. Soc.}\ }\textbf {\bibinfo {volume} {449}},\ \bibinfo {pages} {835} (\bibinfo {year} {2015})},\ \Eprint {http://arxiv.org/abs/1409.3242} {arXiv:1409.3242 [astro-ph.CO]} \BibitemShut {NoStop}%
\bibitem [{\citenamefont {Scolnic}\ \emph {et~al.}(2022)\citenamefont {Scolnic} \emph {et~al.}}]{Scolnic:2021amr}%
  \BibitemOpen
  \bibfield  {author} {\bibinfo {author} {\bibfnamefont {D.}~\bibnamefont {Scolnic}} \emph {et~al.},\ }\href {\doibase 10.3847/1538-4357/ac8b7a} {\bibfield  {journal} {\bibinfo  {journal} {Astrophys. J.}\ }\textbf {\bibinfo {volume} {938}},\ \bibinfo {pages} {113} (\bibinfo {year} {2022})},\ \Eprint {http://arxiv.org/abs/2112.03863} {arXiv:2112.03863 [astro-ph.CO]} \BibitemShut {NoStop}%
\bibitem [{\citenamefont {Peterson}\ \emph {et~al.}(2022)\citenamefont {Peterson} \emph {et~al.}}]{Peterson:2021hel}%
  \BibitemOpen
  \bibfield  {author} {\bibinfo {author} {\bibfnamefont {E.~R.}\ \bibnamefont {Peterson}} \emph {et~al.},\ }\href {\doibase 10.3847/1538-4357/ac4698} {\bibfield  {journal} {\bibinfo  {journal} {Astrophys. J.}\ }\textbf {\bibinfo {volume} {938}},\ \bibinfo {pages} {112} (\bibinfo {year} {2022})},\ \Eprint {http://arxiv.org/abs/2110.03487} {arXiv:2110.03487 [astro-ph.CO]} \BibitemShut {NoStop}%
\bibitem [{\citenamefont {Rubin}\ \emph {et~al.}(2023)\citenamefont {Rubin} \emph {et~al.}}]{Rubin:2023ovl}%
  \BibitemOpen
  \bibfield  {author} {\bibinfo {author} {\bibfnamefont {D.}~\bibnamefont {Rubin}} \emph {et~al.},\ }\href@noop {} {\  (\bibinfo {year} {2023})},\ \Eprint {http://arxiv.org/abs/2311.12098} {arXiv:2311.12098 [astro-ph.CO]} \BibitemShut {NoStop}%
\bibitem [{\citenamefont {Abbott}\ \emph {et~al.}(2024)\citenamefont {Abbott} \emph {et~al.}}]{DES:2024jxu}%
  \BibitemOpen
  \bibfield  {author} {\bibinfo {author} {\bibfnamefont {T.~M.~C.}\ \bibnamefont {Abbott}} \emph {et~al.} (\bibinfo {collaboration} {DES}),\ }\href {\doibase 10.3847/2041-8213/ad6f9f} {\bibfield  {journal} {\bibinfo  {journal} {Astrophys. J. Lett.}\ }\textbf {\bibinfo {volume} {973}},\ \bibinfo {pages} {L14} (\bibinfo {year} {2024})},\ \Eprint {http://arxiv.org/abs/2401.02929} {arXiv:2401.02929 [astro-ph.CO]} \BibitemShut {NoStop}%
\bibitem [{\citenamefont {Akrami}\ \emph {et~al.}(2020)\citenamefont {Akrami} \emph {et~al.}}]{Planck:2020olo}%
  \BibitemOpen
  \bibfield  {author} {\bibinfo {author} {\bibfnamefont {Y.}~\bibnamefont {Akrami}} \emph {et~al.} (\bibinfo {collaboration} {Planck}),\ }\href {\doibase 10.1051/0004-6361/202038073} {\bibfield  {journal} {\bibinfo  {journal} {Astron. Astrophys.}\ }\textbf {\bibinfo {volume} {643}},\ \bibinfo {pages} {A42} (\bibinfo {year} {2020})},\ \Eprint {http://arxiv.org/abs/2007.04997} {arXiv:2007.04997 [astro-ph.CO]} \BibitemShut {NoStop}%
\bibitem [{\citenamefont {Couchot}\ \emph {et~al.}(2017)\citenamefont {Couchot}, \citenamefont {Henrot-Versill\'e}, \citenamefont {Perdereau}, \citenamefont {Plaszczynski}, \citenamefont {Rouill\'e~d'Orfeuil}, \citenamefont {Spinelli},\ and\ \citenamefont {Tristram}}]{Couchot:2016vaq}%
  \BibitemOpen
  \bibfield  {author} {\bibinfo {author} {\bibfnamefont {F.}~\bibnamefont {Couchot}}, \bibinfo {author} {\bibfnamefont {S.}~\bibnamefont {Henrot-Versill\'e}}, \bibinfo {author} {\bibfnamefont {O.}~\bibnamefont {Perdereau}}, \bibinfo {author} {\bibfnamefont {S.}~\bibnamefont {Plaszczynski}}, \bibinfo {author} {\bibfnamefont {B.}~\bibnamefont {Rouill\'e~d'Orfeuil}}, \bibinfo {author} {\bibfnamefont {M.}~\bibnamefont {Spinelli}}, \ and\ \bibinfo {author} {\bibfnamefont {M.}~\bibnamefont {Tristram}},\ }\href {\doibase 10.1051/0004-6361/201629815} {\bibfield  {journal} {\bibinfo  {journal} {Astron. Astrophys.}\ }\textbf {\bibinfo {volume} {602}},\ \bibinfo {pages} {A41} (\bibinfo {year} {2017})},\ \Eprint {http://arxiv.org/abs/1609.09730} {arXiv:1609.09730 [astro-ph.CO]} \BibitemShut {NoStop}%
\bibitem [{\citenamefont {Tristram}\ \emph {et~al.}(2021)\citenamefont {Tristram} \emph {et~al.}}]{Tristram:2020wbi}%
  \BibitemOpen
  \bibfield  {author} {\bibinfo {author} {\bibfnamefont {M.}~\bibnamefont {Tristram}} \emph {et~al.},\ }\href {\doibase 10.1051/0004-6361/202039585} {\bibfield  {journal} {\bibinfo  {journal} {Astron. Astrophys.}\ }\textbf {\bibinfo {volume} {647}},\ \bibinfo {pages} {A128} (\bibinfo {year} {2021})},\ \Eprint {http://arxiv.org/abs/2010.01139} {arXiv:2010.01139 [astro-ph.CO]} \BibitemShut {NoStop}%
\bibitem [{\citenamefont {Aghanim}\ \emph {et~al.}(2020{\natexlab{b}})\citenamefont {Aghanim} \emph {et~al.}}]{Planck:2019nip}%
  \BibitemOpen
  \bibfield  {author} {\bibinfo {author} {\bibfnamefont {N.}~\bibnamefont {Aghanim}} \emph {et~al.} (\bibinfo {collaboration} {Planck}),\ }\href {\doibase 10.1051/0004-6361/201936386} {\bibfield  {journal} {\bibinfo  {journal} {Astron. Astrophys.}\ }\textbf {\bibinfo {volume} {641}},\ \bibinfo {pages} {A5} (\bibinfo {year} {2020}{\natexlab{b}})},\ \Eprint {http://arxiv.org/abs/1907.12875} {arXiv:1907.12875 [astro-ph.CO]} \BibitemShut {NoStop}%
\bibitem [{\citenamefont {Carron}\ \emph {et~al.}(2022)\citenamefont {Carron}, \citenamefont {Mirmelstein},\ and\ \citenamefont {Lewis}}]{Carron:2022eyg}%
  \BibitemOpen
  \bibfield  {author} {\bibinfo {author} {\bibfnamefont {J.}~\bibnamefont {Carron}}, \bibinfo {author} {\bibfnamefont {M.}~\bibnamefont {Mirmelstein}}, \ and\ \bibinfo {author} {\bibfnamefont {A.}~\bibnamefont {Lewis}},\ }\href {\doibase 10.1088/1475-7516/2022/09/039} {\bibfield  {journal} {\bibinfo  {journal} {JCAP}\ }\textbf {\bibinfo {volume} {09}},\ \bibinfo {pages} {039} (\bibinfo {year} {2022})},\ \Eprint {http://arxiv.org/abs/2206.07773} {arXiv:2206.07773 [astro-ph.CO]} \BibitemShut {NoStop}%
\bibitem [{\citenamefont {Madhavacheril}\ \emph {et~al.}(2024)\citenamefont {Madhavacheril} \emph {et~al.}}]{ACT:2023kun}%
  \BibitemOpen
  \bibfield  {author} {\bibinfo {author} {\bibfnamefont {M.~S.}\ \bibnamefont {Madhavacheril}} \emph {et~al.} (\bibinfo {collaboration} {ACT}),\ }\href {\doibase 10.3847/1538-4357/acff5f} {\bibfield  {journal} {\bibinfo  {journal} {Astrophys. J.}\ }\textbf {\bibinfo {volume} {962}},\ \bibinfo {pages} {113} (\bibinfo {year} {2024})},\ \Eprint {http://arxiv.org/abs/2304.05203} {arXiv:2304.05203 [astro-ph.CO]} \BibitemShut {NoStop}%
\bibitem [{\citenamefont {Ade}\ \emph {et~al.}(2021)\citenamefont {Ade} \emph {et~al.}}]{BICEP:2021xfz}%
  \BibitemOpen
  \bibfield  {author} {\bibinfo {author} {\bibfnamefont {P.~A.~R.}\ \bibnamefont {Ade}} \emph {et~al.} (\bibinfo {collaboration} {BICEP, Keck}),\ }\href {\doibase 10.1103/PhysRevLett.127.151301} {\bibfield  {journal} {\bibinfo  {journal} {Phys. Rev. Lett.}\ }\textbf {\bibinfo {volume} {127}},\ \bibinfo {pages} {151301} (\bibinfo {year} {2021})},\ \Eprint {http://arxiv.org/abs/2110.00483} {arXiv:2110.00483 [astro-ph.CO]} \BibitemShut {NoStop}%
\bibitem [{\citenamefont {Di~Valentino}\ \emph {et~al.}(2019)\citenamefont {Di~Valentino}, \citenamefont {Melchiorri},\ and\ \citenamefont {Silk}}]{DiValentino:2019qzk}%
  \BibitemOpen
  \bibfield  {author} {\bibinfo {author} {\bibfnamefont {E.}~\bibnamefont {Di~Valentino}}, \bibinfo {author} {\bibfnamefont {A.}~\bibnamefont {Melchiorri}}, \ and\ \bibinfo {author} {\bibfnamefont {J.}~\bibnamefont {Silk}},\ }\href {\doibase 10.1038/s41550-019-0906-9} {\bibfield  {journal} {\bibinfo  {journal} {Nature Astron.}\ }\textbf {\bibinfo {volume} {4}},\ \bibinfo {pages} {196} (\bibinfo {year} {2019})},\ \Eprint {http://arxiv.org/abs/1911.02087} {arXiv:1911.02087 [astro-ph.CO]} \BibitemShut {NoStop}%
\bibitem [{\citenamefont {Handley}(2021)}]{Handley:2019tkm}%
  \BibitemOpen
  \bibfield  {author} {\bibinfo {author} {\bibfnamefont {W.}~\bibnamefont {Handley}},\ }\href {\doibase 10.1103/PhysRevD.103.L041301} {\bibfield  {journal} {\bibinfo  {journal} {Phys. Rev. D}\ }\textbf {\bibinfo {volume} {103}},\ \bibinfo {pages} {L041301} (\bibinfo {year} {2021})},\ \Eprint {http://arxiv.org/abs/1908.09139} {arXiv:1908.09139 [astro-ph.CO]} \BibitemShut {NoStop}%
\bibitem [{\citenamefont {Efstathiou}\ and\ \citenamefont {Gratton}(2020)}]{Efstathiou:2020wem}%
  \BibitemOpen
  \bibfield  {author} {\bibinfo {author} {\bibfnamefont {G.}~\bibnamefont {Efstathiou}}\ and\ \bibinfo {author} {\bibfnamefont {S.}~\bibnamefont {Gratton}},\ }\href {\doibase 10.1093/mnrasl/slaa093} {\bibfield  {journal} {\bibinfo  {journal} {Mon. Not. Roy. Astron. Soc.}\ }\textbf {\bibinfo {volume} {496}},\ \bibinfo {pages} {L91} (\bibinfo {year} {2020})},\ \Eprint {http://arxiv.org/abs/2002.06892} {arXiv:2002.06892 [astro-ph.CO]} \BibitemShut {NoStop}%
\bibitem [{\citenamefont {Di~Valentino}\ \emph {et~al.}(2021)\citenamefont {Di~Valentino}, \citenamefont {Melchiorri},\ and\ \citenamefont {Silk}}]{DiValentino:2020hov}%
  \BibitemOpen
  \bibfield  {author} {\bibinfo {author} {\bibfnamefont {E.}~\bibnamefont {Di~Valentino}}, \bibinfo {author} {\bibfnamefont {A.}~\bibnamefont {Melchiorri}}, \ and\ \bibinfo {author} {\bibfnamefont {J.}~\bibnamefont {Silk}},\ }\href {\doibase 10.3847/2041-8213/abe1c4} {\bibfield  {journal} {\bibinfo  {journal} {Astrophys. J. Lett.}\ }\textbf {\bibinfo {volume} {908}},\ \bibinfo {pages} {L9} (\bibinfo {year} {2021})},\ \Eprint {http://arxiv.org/abs/2003.04935} {arXiv:2003.04935 [astro-ph.CO]} \BibitemShut {NoStop}%
\bibitem [{\citenamefont {Benisty}\ and\ \citenamefont {Staicova}(2021)}]{Benisty:2020otr}%
  \BibitemOpen
  \bibfield  {author} {\bibinfo {author} {\bibfnamefont {D.}~\bibnamefont {Benisty}}\ and\ \bibinfo {author} {\bibfnamefont {D.}~\bibnamefont {Staicova}},\ }\href {\doibase 10.1051/0004-6361/202039502} {\bibfield  {journal} {\bibinfo  {journal} {Astron. Astrophys.}\ }\textbf {\bibinfo {volume} {647}},\ \bibinfo {pages} {A38} (\bibinfo {year} {2021})},\ \Eprint {http://arxiv.org/abs/2009.10701} {arXiv:2009.10701 [astro-ph.CO]} \BibitemShut {NoStop}%
\bibitem [{\citenamefont {Vagnozzi}\ \emph {et~al.}(2021{\natexlab{a}})\citenamefont {Vagnozzi}, \citenamefont {Di~Valentino}, \citenamefont {Gariazzo}, \citenamefont {Melchiorri}, \citenamefont {Mena},\ and\ \citenamefont {Silk}}]{Vagnozzi:2020rcz}%
  \BibitemOpen
  \bibfield  {author} {\bibinfo {author} {\bibfnamefont {S.}~\bibnamefont {Vagnozzi}}, \bibinfo {author} {\bibfnamefont {E.}~\bibnamefont {Di~Valentino}}, \bibinfo {author} {\bibfnamefont {S.}~\bibnamefont {Gariazzo}}, \bibinfo {author} {\bibfnamefont {A.}~\bibnamefont {Melchiorri}}, \bibinfo {author} {\bibfnamefont {O.}~\bibnamefont {Mena}}, \ and\ \bibinfo {author} {\bibfnamefont {J.}~\bibnamefont {Silk}},\ }\href {\doibase 10.1016/j.dark.2021.100851} {\bibfield  {journal} {\bibinfo  {journal} {Phys. Dark Univ.}\ }\textbf {\bibinfo {volume} {33}},\ \bibinfo {pages} {100851} (\bibinfo {year} {2021}{\natexlab{a}})},\ \Eprint {http://arxiv.org/abs/2010.02230} {arXiv:2010.02230 [astro-ph.CO]} \BibitemShut {NoStop}%
\bibitem [{\citenamefont {Vagnozzi}\ \emph {et~al.}(2021{\natexlab{b}})\citenamefont {Vagnozzi}, \citenamefont {Loeb},\ and\ \citenamefont {Moresco}}]{Vagnozzi:2020dfn}%
  \BibitemOpen
  \bibfield  {author} {\bibinfo {author} {\bibfnamefont {S.}~\bibnamefont {Vagnozzi}}, \bibinfo {author} {\bibfnamefont {A.}~\bibnamefont {Loeb}}, \ and\ \bibinfo {author} {\bibfnamefont {M.}~\bibnamefont {Moresco}},\ }\href {\doibase 10.3847/1538-4357/abd4df} {\bibfield  {journal} {\bibinfo  {journal} {Astrophys. J.}\ }\textbf {\bibinfo {volume} {908}},\ \bibinfo {pages} {84} (\bibinfo {year} {2021}{\natexlab{b}})},\ \Eprint {http://arxiv.org/abs/2011.11645} {arXiv:2011.11645 [astro-ph.CO]} \BibitemShut {NoStop}%
\bibitem [{\citenamefont {Yang}\ \emph {et~al.}(2021)\citenamefont {Yang}, \citenamefont {Pan}, \citenamefont {Di~Valentino}, \citenamefont {Mena},\ and\ \citenamefont {Melchiorri}}]{Yang:2021hxg}%
  \BibitemOpen
  \bibfield  {author} {\bibinfo {author} {\bibfnamefont {W.}~\bibnamefont {Yang}}, \bibinfo {author} {\bibfnamefont {S.}~\bibnamefont {Pan}}, \bibinfo {author} {\bibfnamefont {E.}~\bibnamefont {Di~Valentino}}, \bibinfo {author} {\bibfnamefont {O.}~\bibnamefont {Mena}}, \ and\ \bibinfo {author} {\bibfnamefont {A.}~\bibnamefont {Melchiorri}},\ }\href {\doibase 10.1088/1475-7516/2021/10/008} {\bibfield  {journal} {\bibinfo  {journal} {JCAP}\ }\textbf {\bibinfo {volume} {10}},\ \bibinfo {pages} {008} (\bibinfo {year} {2021})},\ \Eprint {http://arxiv.org/abs/2101.03129} {arXiv:2101.03129 [astro-ph.CO]} \BibitemShut {NoStop}%
\bibitem [{\citenamefont {Cao}\ \emph {et~al.}(2021)\citenamefont {Cao}, \citenamefont {Ryan},\ and\ \citenamefont {Ratra}}]{Cao:2021ldv}%
  \BibitemOpen
  \bibfield  {author} {\bibinfo {author} {\bibfnamefont {S.}~\bibnamefont {Cao}}, \bibinfo {author} {\bibfnamefont {J.}~\bibnamefont {Ryan}}, \ and\ \bibinfo {author} {\bibfnamefont {B.}~\bibnamefont {Ratra}},\ }\href {\doibase 10.1093/mnras/stab942} {\bibfield  {journal} {\bibinfo  {journal} {Mon. Not. Roy. Astron. Soc.}\ }\textbf {\bibinfo {volume} {504}},\ \bibinfo {pages} {300} (\bibinfo {year} {2021})},\ \Eprint {http://arxiv.org/abs/2101.08817} {arXiv:2101.08817 [astro-ph.CO]} \BibitemShut {NoStop}%
\bibitem [{\citenamefont {Gonzalez}\ \emph {et~al.}(2021)\citenamefont {Gonzalez}, \citenamefont {Benetti}, \citenamefont {von Marttens},\ and\ \citenamefont {Alcaniz}}]{Gonzalez:2021ojp}%
  \BibitemOpen
  \bibfield  {author} {\bibinfo {author} {\bibfnamefont {J.~E.}\ \bibnamefont {Gonzalez}}, \bibinfo {author} {\bibfnamefont {M.}~\bibnamefont {Benetti}}, \bibinfo {author} {\bibfnamefont {R.}~\bibnamefont {von Marttens}}, \ and\ \bibinfo {author} {\bibfnamefont {J.}~\bibnamefont {Alcaniz}},\ }\href {\doibase 10.1088/1475-7516/2021/11/060} {\bibfield  {journal} {\bibinfo  {journal} {JCAP}\ }\textbf {\bibinfo {volume} {11}},\ \bibinfo {pages} {060} (\bibinfo {year} {2021})},\ \Eprint {http://arxiv.org/abs/2104.13455} {arXiv:2104.13455 [astro-ph.CO]} \BibitemShut {NoStop}%
\bibitem [{\citenamefont {Dinda}(2022)}]{Dinda:2021ffa}%
  \BibitemOpen
  \bibfield  {author} {\bibinfo {author} {\bibfnamefont {B.~R.}\ \bibnamefont {Dinda}},\ }\href {\doibase 10.1103/PhysRevD.105.063524} {\bibfield  {journal} {\bibinfo  {journal} {Phys. Rev. D}\ }\textbf {\bibinfo {volume} {105}},\ \bibinfo {pages} {063524} (\bibinfo {year} {2022})},\ \Eprint {http://arxiv.org/abs/2106.02963} {arXiv:2106.02963 [astro-ph.CO]} \BibitemShut {NoStop}%
\bibitem [{\citenamefont {Zuckerman}\ and\ \citenamefont {Anchordoqui}(2022)}]{Zuckerman:2021kgm}%
  \BibitemOpen
  \bibfield  {author} {\bibinfo {author} {\bibfnamefont {E.}~\bibnamefont {Zuckerman}}\ and\ \bibinfo {author} {\bibfnamefont {L.~A.}\ \bibnamefont {Anchordoqui}},\ }\href {\doibase 10.1016/j.jheap.2021.10.002} {\bibfield  {journal} {\bibinfo  {journal} {JHEAp}\ }\textbf {\bibinfo {volume} {33}},\ \bibinfo {pages} {10} (\bibinfo {year} {2022})},\ \Eprint {http://arxiv.org/abs/2110.05346} {arXiv:2110.05346 [astro-ph.CO]} \BibitemShut {NoStop}%
\bibitem [{\citenamefont {Bargiacchi}\ \emph {et~al.}(2022)\citenamefont {Bargiacchi}, \citenamefont {Benetti}, \citenamefont {Capozziello}, \citenamefont {Lusso}, \citenamefont {Risaliti},\ and\ \citenamefont {Signorini}}]{Bargiacchi:2021hdp}%
  \BibitemOpen
  \bibfield  {author} {\bibinfo {author} {\bibfnamefont {G.}~\bibnamefont {Bargiacchi}}, \bibinfo {author} {\bibfnamefont {M.}~\bibnamefont {Benetti}}, \bibinfo {author} {\bibfnamefont {S.}~\bibnamefont {Capozziello}}, \bibinfo {author} {\bibfnamefont {E.}~\bibnamefont {Lusso}}, \bibinfo {author} {\bibfnamefont {G.}~\bibnamefont {Risaliti}}, \ and\ \bibinfo {author} {\bibfnamefont {M.}~\bibnamefont {Signorini}},\ }\href {\doibase 10.1093/mnras/stac1941} {\bibfield  {journal} {\bibinfo  {journal} {Mon. Not. Roy. Astron. Soc.}\ }\textbf {\bibinfo {volume} {515}},\ \bibinfo {pages} {1795} (\bibinfo {year} {2022})},\ \Eprint {http://arxiv.org/abs/2111.02420} {arXiv:2111.02420 [astro-ph.CO]} \BibitemShut {NoStop}%
\bibitem [{\citenamefont {Glanville}\ \emph {et~al.}(2022)\citenamefont {Glanville}, \citenamefont {Howlett},\ and\ \citenamefont {Davis}}]{Glanville:2022xes}%
  \BibitemOpen
  \bibfield  {author} {\bibinfo {author} {\bibfnamefont {A.}~\bibnamefont {Glanville}}, \bibinfo {author} {\bibfnamefont {C.}~\bibnamefont {Howlett}}, \ and\ \bibinfo {author} {\bibfnamefont {T.~M.}\ \bibnamefont {Davis}},\ }\href {\doibase 10.1093/mnras/stac2891} {\bibfield  {journal} {\bibinfo  {journal} {Mon. Not. Roy. Astron. Soc.}\ }\textbf {\bibinfo {volume} {517}},\ \bibinfo {pages} {3087} (\bibinfo {year} {2022})},\ \Eprint {http://arxiv.org/abs/2205.05892} {arXiv:2205.05892 [astro-ph.CO]} \BibitemShut {NoStop}%
\bibitem [{\citenamefont {Bel}\ \emph {et~al.}(2022)\citenamefont {Bel}, \citenamefont {Larena}, \citenamefont {Maartens}, \citenamefont {Marinoni},\ and\ \citenamefont {Perenon}}]{Bel:2022iuf}%
  \BibitemOpen
  \bibfield  {author} {\bibinfo {author} {\bibfnamefont {J.}~\bibnamefont {Bel}}, \bibinfo {author} {\bibfnamefont {J.}~\bibnamefont {Larena}}, \bibinfo {author} {\bibfnamefont {R.}~\bibnamefont {Maartens}}, \bibinfo {author} {\bibfnamefont {C.}~\bibnamefont {Marinoni}}, \ and\ \bibinfo {author} {\bibfnamefont {L.}~\bibnamefont {Perenon}},\ }\href {\doibase 10.1088/1475-7516/2022/09/076} {\bibfield  {journal} {\bibinfo  {journal} {JCAP}\ }\textbf {\bibinfo {volume} {09}},\ \bibinfo {pages} {076} (\bibinfo {year} {2022})},\ \Eprint {http://arxiv.org/abs/2206.03059} {arXiv:2206.03059 [astro-ph.CO]} \BibitemShut {NoStop}%
\bibitem [{\citenamefont {Yang}\ \emph {et~al.}(2023)\citenamefont {Yang}, \citenamefont {Giar\`e}, \citenamefont {Pan}, \citenamefont {Di~Valentino}, \citenamefont {Melchiorri},\ and\ \citenamefont {Silk}}]{Yang:2022kho}%
  \BibitemOpen
  \bibfield  {author} {\bibinfo {author} {\bibfnamefont {W.}~\bibnamefont {Yang}}, \bibinfo {author} {\bibfnamefont {W.}~\bibnamefont {Giar\`e}}, \bibinfo {author} {\bibfnamefont {S.}~\bibnamefont {Pan}}, \bibinfo {author} {\bibfnamefont {E.}~\bibnamefont {Di~Valentino}}, \bibinfo {author} {\bibfnamefont {A.}~\bibnamefont {Melchiorri}}, \ and\ \bibinfo {author} {\bibfnamefont {J.}~\bibnamefont {Silk}},\ }\href {\doibase 10.1103/PhysRevD.107.063509} {\bibfield  {journal} {\bibinfo  {journal} {Phys. Rev. D}\ }\textbf {\bibinfo {volume} {107}},\ \bibinfo {pages} {063509} (\bibinfo {year} {2023})},\ \Eprint {http://arxiv.org/abs/2210.09865} {arXiv:2210.09865 [astro-ph.CO]} \BibitemShut {NoStop}%
\bibitem [{\citenamefont {Stevens}\ \emph {et~al.}(2023)\citenamefont {Stevens}, \citenamefont {Khoraminezhad},\ and\ \citenamefont {Saito}}]{Stevens:2022evv}%
  \BibitemOpen
  \bibfield  {author} {\bibinfo {author} {\bibfnamefont {J.}~\bibnamefont {Stevens}}, \bibinfo {author} {\bibfnamefont {H.}~\bibnamefont {Khoraminezhad}}, \ and\ \bibinfo {author} {\bibfnamefont {S.}~\bibnamefont {Saito}},\ }\href {\doibase 10.1088/1475-7516/2023/07/046} {\bibfield  {journal} {\bibinfo  {journal} {JCAP}\ }\textbf {\bibinfo {volume} {07}},\ \bibinfo {pages} {046} (\bibinfo {year} {2023})},\ \Eprint {http://arxiv.org/abs/2212.09804} {arXiv:2212.09804 [astro-ph.CO]} \BibitemShut {NoStop}%
\bibitem [{\citenamefont {Favale}\ \emph {et~al.}(2023)\citenamefont {Favale}, \citenamefont {G\'omez-Valent},\ and\ \citenamefont {Migliaccio}}]{Favale:2023lnp}%
  \BibitemOpen
  \bibfield  {author} {\bibinfo {author} {\bibfnamefont {A.}~\bibnamefont {Favale}}, \bibinfo {author} {\bibfnamefont {A.}~\bibnamefont {G\'omez-Valent}}, \ and\ \bibinfo {author} {\bibfnamefont {M.}~\bibnamefont {Migliaccio}},\ }\href {\doibase 10.1093/mnras/stad1621} {\bibfield  {journal} {\bibinfo  {journal} {Mon. Not. Roy. Astron. Soc.}\ }\textbf {\bibinfo {volume} {523}},\ \bibinfo {pages} {3406} (\bibinfo {year} {2023})},\ \Eprint {http://arxiv.org/abs/2301.09591} {arXiv:2301.09591 [astro-ph.CO]} \BibitemShut {NoStop}%
\bibitem [{\citenamefont {Qi}\ \emph {et~al.}(2023)\citenamefont {Qi}, \citenamefont {Meng}, \citenamefont {Zhang},\ and\ \citenamefont {Zhang}}]{Qi:2023oxv}%
  \BibitemOpen
  \bibfield  {author} {\bibinfo {author} {\bibfnamefont {J.-Z.}\ \bibnamefont {Qi}}, \bibinfo {author} {\bibfnamefont {P.}~\bibnamefont {Meng}}, \bibinfo {author} {\bibfnamefont {J.-F.}\ \bibnamefont {Zhang}}, \ and\ \bibinfo {author} {\bibfnamefont {X.}~\bibnamefont {Zhang}},\ }\href {\doibase 10.1103/PhysRevD.108.063522} {\bibfield  {journal} {\bibinfo  {journal} {Phys. Rev. D}\ }\textbf {\bibinfo {volume} {108}},\ \bibinfo {pages} {063522} (\bibinfo {year} {2023})},\ \Eprint {http://arxiv.org/abs/2302.08889} {arXiv:2302.08889 [astro-ph.CO]} \BibitemShut {NoStop}%
\bibitem [{\citenamefont {Chevallier}\ and\ \citenamefont {Polarski}(2001)}]{Chevallier:2000qy}%
  \BibitemOpen
  \bibfield  {author} {\bibinfo {author} {\bibfnamefont {M.}~\bibnamefont {Chevallier}}\ and\ \bibinfo {author} {\bibfnamefont {D.}~\bibnamefont {Polarski}},\ }\href {\doibase 10.1142/S0218271801000822} {\bibfield  {journal} {\bibinfo  {journal} {Int. J. Mod. Phys. D}\ }\textbf {\bibinfo {volume} {10}},\ \bibinfo {pages} {213} (\bibinfo {year} {2001})},\ \Eprint {http://arxiv.org/abs/gr-qc/0009008} {arXiv:gr-qc/0009008} \BibitemShut {NoStop}%
\bibitem [{\citenamefont {Linder}(2003)}]{Linder:2002et}%
  \BibitemOpen
  \bibfield  {author} {\bibinfo {author} {\bibfnamefont {E.~V.}\ \bibnamefont {Linder}},\ }\href {\doibase 10.1103/PhysRevLett.90.091301} {\bibfield  {journal} {\bibinfo  {journal} {Phys. Rev. Lett.}\ }\textbf {\bibinfo {volume} {90}},\ \bibinfo {pages} {091301} (\bibinfo {year} {2003})},\ \Eprint {http://arxiv.org/abs/astro-ph/0208512} {arXiv:astro-ph/0208512} \BibitemShut {NoStop}%
\bibitem [{\citenamefont {Vagnozzi}\ \emph {et~al.}(2018)\citenamefont {Vagnozzi}, \citenamefont {Dhawan}, \citenamefont {Gerbino}, \citenamefont {Freese}, \citenamefont {Goobar},\ and\ \citenamefont {Mena}}]{Vagnozzi:2018jhn}%
  \BibitemOpen
  \bibfield  {author} {\bibinfo {author} {\bibfnamefont {S.}~\bibnamefont {Vagnozzi}}, \bibinfo {author} {\bibfnamefont {S.}~\bibnamefont {Dhawan}}, \bibinfo {author} {\bibfnamefont {M.}~\bibnamefont {Gerbino}}, \bibinfo {author} {\bibfnamefont {K.}~\bibnamefont {Freese}}, \bibinfo {author} {\bibfnamefont {A.}~\bibnamefont {Goobar}}, \ and\ \bibinfo {author} {\bibfnamefont {O.}~\bibnamefont {Mena}},\ }\href {\doibase 10.1103/PhysRevD.98.083501} {\bibfield  {journal} {\bibinfo  {journal} {Phys. Rev. D}\ }\textbf {\bibinfo {volume} {98}},\ \bibinfo {pages} {083501} (\bibinfo {year} {2018})},\ \Eprint {http://arxiv.org/abs/1801.08553} {arXiv:1801.08553 [astro-ph.CO]} \BibitemShut {NoStop}%
\bibitem [{\citenamefont {Akarsu}\ \emph {et~al.}(2021)\citenamefont {Akarsu}, \citenamefont {Kumar}, \citenamefont {\"Oz\"ulker},\ and\ \citenamefont {Vazquez}}]{Akarsu:2021fol}%
  \BibitemOpen
  \bibfield  {author} {\bibinfo {author} {\bibfnamefont {O.}~\bibnamefont {Akarsu}}, \bibinfo {author} {\bibfnamefont {S.}~\bibnamefont {Kumar}}, \bibinfo {author} {\bibfnamefont {E.}~\bibnamefont {\"Oz\"ulker}}, \ and\ \bibinfo {author} {\bibfnamefont {J.~A.}\ \bibnamefont {Vazquez}},\ }\href {\doibase 10.1103/PhysRevD.104.123512} {\bibfield  {journal} {\bibinfo  {journal} {Phys. Rev. D}\ }\textbf {\bibinfo {volume} {104}},\ \bibinfo {pages} {123512} (\bibinfo {year} {2021})},\ \Eprint {http://arxiv.org/abs/2108.09239} {arXiv:2108.09239 [astro-ph.CO]} \BibitemShut {NoStop}%
\bibitem [{\citenamefont {Akarsu}\ \emph {et~al.}(2023)\citenamefont {Akarsu}, \citenamefont {Kumar}, \citenamefont {\"Oz\"ulker}, \citenamefont {Vazquez},\ and\ \citenamefont {Yadav}}]{Akarsu:2022typ}%
  \BibitemOpen
  \bibfield  {author} {\bibinfo {author} {\bibfnamefont {O.}~\bibnamefont {Akarsu}}, \bibinfo {author} {\bibfnamefont {S.}~\bibnamefont {Kumar}}, \bibinfo {author} {\bibfnamefont {E.}~\bibnamefont {\"Oz\"ulker}}, \bibinfo {author} {\bibfnamefont {J.~A.}\ \bibnamefont {Vazquez}}, \ and\ \bibinfo {author} {\bibfnamefont {A.}~\bibnamefont {Yadav}},\ }\href {\doibase 10.1103/PhysRevD.108.023513} {\bibfield  {journal} {\bibinfo  {journal} {Phys. Rev. D}\ }\textbf {\bibinfo {volume} {108}},\ \bibinfo {pages} {023513} (\bibinfo {year} {2023})},\ \Eprint {http://arxiv.org/abs/2211.05742} {arXiv:2211.05742 [astro-ph.CO]} \BibitemShut {NoStop}%
\bibitem [{\citenamefont {Toda}\ \emph {et~al.}(2024)\citenamefont {Toda}, \citenamefont {Giar\`e}, \citenamefont {\"Oz\"ulker}, \citenamefont {Di~Valentino},\ and\ \citenamefont {Vagnozzi}}]{Toda:2024ncp}%
  \BibitemOpen
  \bibfield  {author} {\bibinfo {author} {\bibfnamefont {Y.}~\bibnamefont {Toda}}, \bibinfo {author} {\bibfnamefont {W.}~\bibnamefont {Giar\`e}}, \bibinfo {author} {\bibfnamefont {E.}~\bibnamefont {\"Oz\"ulker}}, \bibinfo {author} {\bibfnamefont {E.}~\bibnamefont {Di~Valentino}}, \ and\ \bibinfo {author} {\bibfnamefont {S.}~\bibnamefont {Vagnozzi}},\ }\href {\doibase 10.1016/j.dark.2024.101676} {\bibfield  {journal} {\bibinfo  {journal} {Phys. Dark Univ.}\ }\textbf {\bibinfo {volume} {46}},\ \bibinfo {pages} {101676} (\bibinfo {year} {2024})},\ \Eprint {http://arxiv.org/abs/2407.01173} {arXiv:2407.01173 [astro-ph.CO]} \BibitemShut {NoStop}%
\bibitem [{\citenamefont {Lewis}\ \emph {et~al.}(2000)\citenamefont {Lewis}, \citenamefont {Challinor},\ and\ \citenamefont {Lasenby}}]{Lewis:1999bs}%
  \BibitemOpen
  \bibfield  {author} {\bibinfo {author} {\bibfnamefont {A.}~\bibnamefont {Lewis}}, \bibinfo {author} {\bibfnamefont {A.}~\bibnamefont {Challinor}}, \ and\ \bibinfo {author} {\bibfnamefont {A.}~\bibnamefont {Lasenby}},\ }\href {\doibase 10.1086/309179} {\bibfield  {journal} {\bibinfo  {journal} {\apj}\ }\textbf {\bibinfo {volume} {538}},\ \bibinfo {pages} {473} (\bibinfo {year} {2000})},\ \Eprint {http://arxiv.org/abs/astro-ph/9911177} {arXiv:astro-ph/9911177 [astro-ph]} \BibitemShut {NoStop}%
\bibitem [{\citenamefont {Torrado}\ and\ \citenamefont {Lewis}(2021)}]{Torrado:2020dgo}%
  \BibitemOpen
  \bibfield  {author} {\bibinfo {author} {\bibfnamefont {J.}~\bibnamefont {Torrado}}\ and\ \bibinfo {author} {\bibfnamefont {A.}~\bibnamefont {Lewis}},\ }\href {\doibase 10.1088/1475-7516/2021/05/057} {\bibfield  {journal} {\bibinfo  {journal} {JCAP}\ }\textbf {\bibinfo {volume} {05}},\ \bibinfo {pages} {057} (\bibinfo {year} {2021})},\ \Eprint {http://arxiv.org/abs/2005.05290} {arXiv:2005.05290 [astro-ph.IM]} \BibitemShut {NoStop}%
\bibitem [{\citenamefont {Gelman}\ and\ \citenamefont {Rubin}(1992)}]{Gelman:1992zz}%
  \BibitemOpen
  \bibfield  {author} {\bibinfo {author} {\bibfnamefont {A.}~\bibnamefont {Gelman}}\ and\ \bibinfo {author} {\bibfnamefont {D.~B.}\ \bibnamefont {Rubin}},\ }\href {\doibase 10.1214/ss/1177011136} {\bibfield  {journal} {\bibinfo  {journal} {Statist. Sci.}\ }\textbf {\bibinfo {volume} {7}},\ \bibinfo {pages} {457} (\bibinfo {year} {1992})}\BibitemShut {NoStop}%
\bibitem [{\citenamefont {Lewis}(2019)}]{Lewis:2019xzd}%
  \BibitemOpen
  \bibfield  {author} {\bibinfo {author} {\bibfnamefont {A.}~\bibnamefont {Lewis}},\ }\href@noop {} {\  (\bibinfo {year} {2019})},\ \Eprint {http://arxiv.org/abs/1910.13970} {arXiv:1910.13970 [astro-ph.IM]} \BibitemShut {NoStop}%
\bibitem [{\citenamefont {Hou}\ \emph {et~al.}(2013)\citenamefont {Hou}, \citenamefont {Keisler}, \citenamefont {Knox}, \citenamefont {Millea},\ and\ \citenamefont {Reichardt}}]{Hou:2011ec}%
  \BibitemOpen
  \bibfield  {author} {\bibinfo {author} {\bibfnamefont {Z.}~\bibnamefont {Hou}}, \bibinfo {author} {\bibfnamefont {R.}~\bibnamefont {Keisler}}, \bibinfo {author} {\bibfnamefont {L.}~\bibnamefont {Knox}}, \bibinfo {author} {\bibfnamefont {M.}~\bibnamefont {Millea}}, \ and\ \bibinfo {author} {\bibfnamefont {C.}~\bibnamefont {Reichardt}},\ }\href {\doibase 10.1103/PhysRevD.87.083008} {\bibfield  {journal} {\bibinfo  {journal} {Phys. Rev. D}\ }\textbf {\bibinfo {volume} {87}},\ \bibinfo {pages} {083008} (\bibinfo {year} {2013})},\ \Eprint {http://arxiv.org/abs/1104.2333} {arXiv:1104.2333 [astro-ph.CO]} \BibitemShut {NoStop}%
\bibitem [{\citenamefont {Kable}\ \emph {et~al.}(2020)\citenamefont {Kable}, \citenamefont {Addison},\ and\ \citenamefont {Bennett}}]{Kable:2020hcw}%
  \BibitemOpen
  \bibfield  {author} {\bibinfo {author} {\bibfnamefont {J.~A.}\ \bibnamefont {Kable}}, \bibinfo {author} {\bibfnamefont {G.~E.}\ \bibnamefont {Addison}}, \ and\ \bibinfo {author} {\bibfnamefont {C.~L.}\ \bibnamefont {Bennett}},\ }\href {\doibase 10.3847/1538-4357/abc4e7} {\bibfield  {journal} {\bibinfo  {journal} {Astrophys. J.}\ }\textbf {\bibinfo {volume} {905}},\ \bibinfo {pages} {164} (\bibinfo {year} {2020})},\ \Eprint {http://arxiv.org/abs/2008.01785} {arXiv:2008.01785 [astro-ph.CO]} \BibitemShut {NoStop}%
\bibitem [{\citenamefont {Vagnozzi}(2021)}]{Vagnozzi:2021gjh}%
  \BibitemOpen
  \bibfield  {author} {\bibinfo {author} {\bibfnamefont {S.}~\bibnamefont {Vagnozzi}},\ }\href {\doibase 10.1103/PhysRevD.104.063524} {\bibfield  {journal} {\bibinfo  {journal} {Phys. Rev. D}\ }\textbf {\bibinfo {volume} {104}},\ \bibinfo {pages} {063524} (\bibinfo {year} {2021})},\ \Eprint {http://arxiv.org/abs/2105.10425} {arXiv:2105.10425 [astro-ph.CO]} \BibitemShut {NoStop}%
\bibitem [{\citenamefont {Poulin}\ \emph {et~al.}(2024)\citenamefont {Poulin}, \citenamefont {Smith}, \citenamefont {Calder\'on},\ and\ \citenamefont {Simon}}]{Poulin:2024ken}%
  \BibitemOpen
  \bibfield  {author} {\bibinfo {author} {\bibfnamefont {V.}~\bibnamefont {Poulin}}, \bibinfo {author} {\bibfnamefont {T.~L.}\ \bibnamefont {Smith}}, \bibinfo {author} {\bibfnamefont {R.}~\bibnamefont {Calder\'on}}, \ and\ \bibinfo {author} {\bibfnamefont {T.}~\bibnamefont {Simon}},\ }\href@noop {} {\  (\bibinfo {year} {2024})},\ \Eprint {http://arxiv.org/abs/2407.18292} {arXiv:2407.18292 [astro-ph.CO]} \BibitemShut {NoStop}%
\bibitem [{\citenamefont {Wang}(2024)}]{Wang:2024rjd}%
  \BibitemOpen
  \bibfield  {author} {\bibinfo {author} {\bibfnamefont {D.}~\bibnamefont {Wang}},\ }\href@noop {} {\  (\bibinfo {year} {2024})},\ \Eprint {http://arxiv.org/abs/2404.13833} {arXiv:2404.13833 [astro-ph.CO]} \BibitemShut {NoStop}%
\bibitem [{\citenamefont {Pogosian}\ \emph {et~al.}(2024)\citenamefont {Pogosian}, \citenamefont {Zhao},\ and\ \citenamefont {Jedamzik}}]{Pogosian:2024ykm}%
  \BibitemOpen
  \bibfield  {author} {\bibinfo {author} {\bibfnamefont {L.}~\bibnamefont {Pogosian}}, \bibinfo {author} {\bibfnamefont {G.-B.}\ \bibnamefont {Zhao}}, \ and\ \bibinfo {author} {\bibfnamefont {K.}~\bibnamefont {Jedamzik}},\ }\href {\doibase 10.3847/2041-8213/ad7507} {\bibfield  {journal} {\bibinfo  {journal} {Astrophys. J. Lett.}\ }\textbf {\bibinfo {volume} {973}},\ \bibinfo {pages} {L13} (\bibinfo {year} {2024})},\ \Eprint {http://arxiv.org/abs/2405.20306} {arXiv:2405.20306 [astro-ph.CO]} \BibitemShut {NoStop}%
\bibitem [{\citenamefont {Giar\`e}(2024{\natexlab{b}})}]{Giare:2024akf}%
  \BibitemOpen
  \bibfield  {author} {\bibinfo {author} {\bibfnamefont {W.}~\bibnamefont {Giar\`e}},\ }\href {\doibase 10.1103/PhysRevD.109.123545} {\bibfield  {journal} {\bibinfo  {journal} {Phys. Rev. D}\ }\textbf {\bibinfo {volume} {109}},\ \bibinfo {pages} {123545} (\bibinfo {year} {2024}{\natexlab{b}})},\ \Eprint {http://arxiv.org/abs/2404.12779} {arXiv:2404.12779 [astro-ph.CO]} \BibitemShut {NoStop}%
\bibitem [{\citenamefont {Poulin}\ \emph {et~al.}(2023)\citenamefont {Poulin}, \citenamefont {Smith},\ and\ \citenamefont {Karwal}}]{Poulin:2023lkg}%
  \BibitemOpen
  \bibfield  {author} {\bibinfo {author} {\bibfnamefont {V.}~\bibnamefont {Poulin}}, \bibinfo {author} {\bibfnamefont {T.~L.}\ \bibnamefont {Smith}}, \ and\ \bibinfo {author} {\bibfnamefont {T.}~\bibnamefont {Karwal}},\ }\href {\doibase 10.1016/j.dark.2023.101348} {\bibfield  {journal} {\bibinfo  {journal} {Phys. Dark Univ.}\ }\textbf {\bibinfo {volume} {42}},\ \bibinfo {pages} {101348} (\bibinfo {year} {2023})},\ \Eprint {http://arxiv.org/abs/2302.09032} {arXiv:2302.09032 [astro-ph.CO]} \BibitemShut {NoStop}%
\bibitem [{\citenamefont {Giar\`e}\ \emph {et~al.}(2023)\citenamefont {Giar\`e}, \citenamefont {Renzi}, \citenamefont {Mena}, \citenamefont {Di~Valentino},\ and\ \citenamefont {Melchiorri}}]{Giare:2022rvg}%
  \BibitemOpen
  \bibfield  {author} {\bibinfo {author} {\bibfnamefont {W.}~\bibnamefont {Giar\`e}}, \bibinfo {author} {\bibfnamefont {F.}~\bibnamefont {Renzi}}, \bibinfo {author} {\bibfnamefont {O.}~\bibnamefont {Mena}}, \bibinfo {author} {\bibfnamefont {E.}~\bibnamefont {Di~Valentino}}, \ and\ \bibinfo {author} {\bibfnamefont {A.}~\bibnamefont {Melchiorri}},\ }\href {\doibase 10.1093/mnras/stad724} {\bibfield  {journal} {\bibinfo  {journal} {Mon. Not. Roy. Astron. Soc.}\ }\textbf {\bibinfo {volume} {521}},\ \bibinfo {pages} {2911} (\bibinfo {year} {2023})},\ \Eprint {http://arxiv.org/abs/2210.09018} {arXiv:2210.09018 [astro-ph.CO]} \BibitemShut {NoStop}%
\bibitem [{\citenamefont {Jiang}\ \emph {et~al.}(2023)\citenamefont {Jiang}, \citenamefont {Ye},\ and\ \citenamefont {Piao}}]{Jiang:2022qlj}%
  \BibitemOpen
  \bibfield  {author} {\bibinfo {author} {\bibfnamefont {J.-Q.}\ \bibnamefont {Jiang}}, \bibinfo {author} {\bibfnamefont {G.}~\bibnamefont {Ye}}, \ and\ \bibinfo {author} {\bibfnamefont {Y.-S.}\ \bibnamefont {Piao}},\ }\href {\doibase 10.1093/mnrasl/slad137} {\bibfield  {journal} {\bibinfo  {journal} {Mon. Not. Roy. Astron. Soc.}\ }\textbf {\bibinfo {volume} {527}},\ \bibinfo {pages} {L54} (\bibinfo {year} {2023})},\ \Eprint {http://arxiv.org/abs/2210.06125} {arXiv:2210.06125 [astro-ph.CO]} \BibitemShut {NoStop}%
\bibitem [{\citenamefont {Ye}\ \emph {et~al.}(2022)\citenamefont {Ye}, \citenamefont {Jiang},\ and\ \citenamefont {Piao}}]{Ye:2022efx}%
  \BibitemOpen
  \bibfield  {author} {\bibinfo {author} {\bibfnamefont {G.}~\bibnamefont {Ye}}, \bibinfo {author} {\bibfnamefont {J.-Q.}\ \bibnamefont {Jiang}}, \ and\ \bibinfo {author} {\bibfnamefont {Y.-S.}\ \bibnamefont {Piao}},\ }\href {\doibase 10.1103/PhysRevD.106.103528} {\bibfield  {journal} {\bibinfo  {journal} {Phys. Rev. D}\ }\textbf {\bibinfo {volume} {106}},\ \bibinfo {pages} {103528} (\bibinfo {year} {2022})},\ \Eprint {http://arxiv.org/abs/2205.02478} {arXiv:2205.02478 [astro-ph.CO]} \BibitemShut {NoStop}%
\bibitem [{\citenamefont {Jiang}\ and\ \citenamefont {Piao}(2022)}]{Jiang:2022uyg}%
  \BibitemOpen
  \bibfield  {author} {\bibinfo {author} {\bibfnamefont {J.-Q.}\ \bibnamefont {Jiang}}\ and\ \bibinfo {author} {\bibfnamefont {Y.-S.}\ \bibnamefont {Piao}},\ }\href {\doibase 10.1103/PhysRevD.105.103514} {\bibfield  {journal} {\bibinfo  {journal} {Phys. Rev. D}\ }\textbf {\bibinfo {volume} {105}},\ \bibinfo {pages} {103514} (\bibinfo {year} {2022})},\ \Eprint {http://arxiv.org/abs/2202.13379} {arXiv:2202.13379 [astro-ph.CO]} \BibitemShut {NoStop}%
\end{thebibliography}%
\end{document}